\documentclass[twocolumn]{aastex63}

\usepackage{color}

\received{}
\revised{}
\accepted{}

\submitjournal{ApJ}

\shorttitle{Multi-wavelength View of Type IIn Supernovae}
\shortauthors{Tsuna, Kashiyama, Shigeyama}

\begin{document}

\title{
Multi-wavelength View of the Type IIn Zoo: 
Optical to X-ray Emission Model of Interaction-Powered Supernovae}

\correspondingauthor{Daichi Tsuna}
\email{tsuna@resceu.s.u-tokyo.ac.jp}

 \author{Daichi Tsuna}
 \affiliation{Research Center for the Early Universe (RESCEU), School of Science, The University of Tokyo, 7-3-1 Hongo, Bunkyo-ku, Tokyo 113-0033, Japan}
 \affiliation{Department of Physics, School of Science, The University of Tokyo, 7-3-1 Hongo, Bunkyo-ku, Tokyo 113-0033, Japan}

\author{Kazumi Kashiyama}
 \affiliation{Research Center for the Early Universe (RESCEU), School of Science, The University of Tokyo, 7-3-1 Hongo, Bunkyo-ku, Tokyo 113-0033, Japan}
 \affiliation{Department of Physics, School of Science, The University of Tokyo, 7-3-1 Hongo, Bunkyo-ku, Tokyo 113-0033, Japan}

 \author{Toshikazu Shigeyama}
 \affiliation{Research Center for the Early Universe (RESCEU), School of Science, The University of Tokyo, 7-3-1 Hongo, Bunkyo-ku, Tokyo 113-0033, Japan}
 \affiliation{Department of Astronomy, School of Science, The University of Tokyo, 7-3-1 Hongo, Bunkyo-ku, Tokyo 113-0033, Japan}

\begin{abstract}
Transients powered by interaction with the circumstellar medium (CSM) are often observed in wavelengths other than optical, and multi-wavelength modelling can be important when inferring the properties of the explosion and CSM, or for distinguishing from other powering mechanisms.
We develop a model calculating time dependent emission spectrum of interaction-powered transients. 
We solve energy equations of electron-proton plasma in the shocked SN ejecta and CSM and a radiation transfer equation out to the outer edge of the CSM, incorporating the collisional relaxation and the comptonization of the bremsstrahlung radiation.
We compare our model to observations of Type IIn supernovae covering frequency ranges from optical to X-rays. For SN 2010jl the observed optical and X-ray light curves can be consistently explained if clumpy or asymmetric structure in the CSM is assumed, in agreement with previous studies. For SN 2014C our model successfully reproduces the X-ray bremsstrahlung component and the emergence of H$\alpha$ emission at 400 days after explosion. Finally we find a parameter space where the supernova is extremely X-ray bright, reaching $10^{43}$--$10^{44}\ {\rm erg\ s^{-1}}$ for up to $100$ days. Such X-ray transients are likely detectable with all-sky surveys by e.g. eROSITA.
\end{abstract}


\section{Introduction}
Massive stars with initial masses of $>8M_\odot$ are considered to end their lives as supernovae (SNe). Among the diverse photometric and spectroscopic appearances, there have been a growing number of transients that show signs of interaction between the SN ejecta and a pre-existing dense circumstellar medium (CSM).

Type IIn SNe \citep{Schlegel90,Filippenko97}, which display narrow ($10$--$1000\ {\rm km\ s^{-1}}$) hydrogen emission lines in the optical spectra, are the representative for this kind of transients. Such emission lines are attributed to the existence of a slowly moving dense CSM (e.g. \citealt{Chugai91}). The dense CSM also efficiently dissipates the kinetic energy of the SN ejecta (e.g. \citealt{Grasberg86}), which are considered to power the light curve at least in the early phases. In cases where the CSM is dense, the shock breakout from the progenitor itself is delayed and prolonged (e.g. \citealt{Chevalier11,Forster18}).

The collision of SN ejecta with the CSM leads to the formation of shock waves. The shock waves, propagating at a velocity of $10^3$--$10^4\ {\rm km\ s^{-1}}$, heat the upstream gas to temperatures of $1$--$100$ keV. By radiative cooling this plasma emits copious X-ray photons, which are reprocessed in the shocked region and the CSM. X-ray emission is detected in some Type IIn SNe, e.g. in SN 2005ip \citep{Katsuda14}, 2005kd \citep{Dwarkadas16,Katsuda16}, 2006jd \citep{Chandra12_2006jd,Katsuda16}, and 2010jl \citep{Ofek14,Chandra15}. The X-ray luminosity from these SNe ranges from $10^{40}$ to $10^{42}\ {\rm erg\ s^{-1}}$, much brighter than those in the other classes of supernovae \citep{Chandra18}.

Observations in soft and hard X-rays enable us to extract the column depth of neutral hydrogen ahead of the shock. This was done in e.g. SN 2010jl \citep{Chandra15, Katsuda16}, which showed an evolving column density by orders of magnitude over the timescale of a few years. With the natural assumption that the column density is dominated by the CSM, this demonstrates that simultaneous observations in X-rays along with optical can be important for tracing the activity of massive stars in the end of their lives.
    
There are previous works on modelling X-ray emission from interaction-powered transients upon and after shock breakout \citep{Katz10,Balberg11,Chevalier12, Svirski12,Pan13}. While these works enable us to obtain a rough estimate of the X-ray emission, no previous work has been done in obtaining a long-term broadband spectrum that can realize detailed comparison with observations.

In this work we attempt to model a multi-wavelength emission of interaction-powered transients from optical to X-rays. In our model we take into account the basic radiative processes governing inside the shocked region as well as the CSM. We find that the density (mass-loss rate) of the CSM is crucial in shaping the optical--X-ray spectra, and that the X-ray emission can be quite bright and detectable when the CSM ahead of the shock front becomes transparent.

This paper is constructed as follows. We describe our emission model in Section \ref{sec:Model}. We present the resulting spectra and optical/X-ray light curves in Section \ref{sec:results}, and compare our model with observations of Type IIn SNe 2010jl and 2014C in Section \ref{sec:comparison}. We list some caveats of our model in Section \ref{sec:caveats}, and conclude in Section \ref{sec:conclusion}.

\section{Model}
\label{sec:Model}
\begin{figure*}[t]
     \centering
     \includegraphics[width=\linewidth]{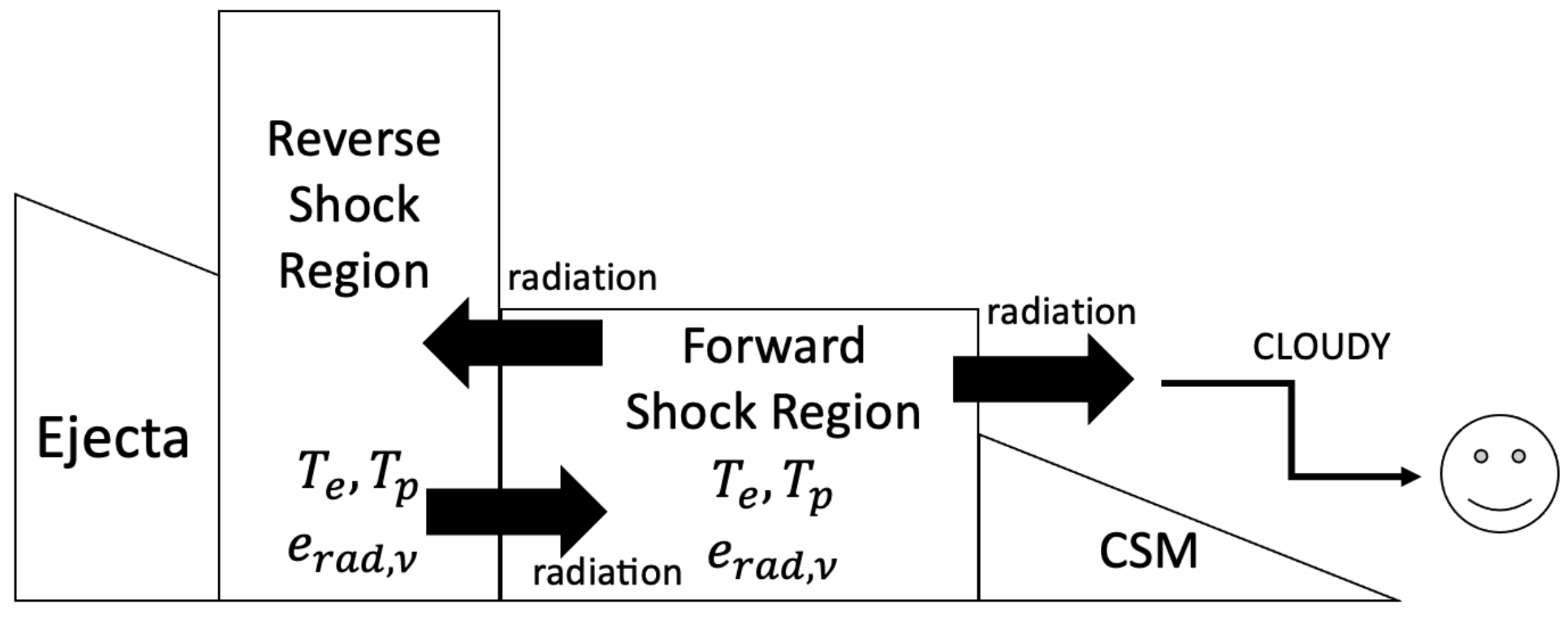}
     \caption{Schematic figure of our emission model (not to scale). For each region, the evolution of electron and proton temperatures and radiation from the shock to the contact surface are solved, with the parameters governing the dynamics (density, velocity, pressure) assumed to be one-zone. For the forward shock region, half of the escaping radiation will be fed into the reverse shock region and the other half escapes, while for the reverse shock region all of the escaping radiation will be fed into the forward shock region. The escaping radiation towards the observer is post-processed with the CLOUDY photoionization code \citep{Ferland17}.}
     \label{fig:schematic}
\end{figure*}

We first outline the basic idea of our emission model, next summarize the basic equations, and then describe how to solve them.
A schematic view of our modelling is shown in Figure \ref{fig:schematic}. 

\subsection{Overview}
\label{sec:summary}

We consider a core-collapse SN occurring in a dense CSM. 
When the SN ejecta collide with the CSM, two shocks are formed; a forward shock (FS) separating the shocked and unshocked CSM and a reverse shock (RS) separating the shocked and unshocked SN ejecta \citep{Chevalier82}. 
The interface of the shocked CSM and SN ejecta is called a contact discontinuity (see Fig. \ref{fig:schematic}). 
We consider the case where homologously expanding cold SN ejecta with a power-law density profile ($\rho_{\rm ej}\propto v^{-n}$ with $n>5$) run through a CSM of a wind density profile
\begin{eqnarray}
    \rho_w = \frac{\dot{M}}{4\pi v_wr^2}\sim 5\times 10^{-15}{\rm g\ cm^{-3}} \left[\frac{\dot{M}/v_w}{(\dot{M}/v_w)_*}\right]\left(\frac{r}{10^{15}{\rm cm}}\right)^{-2},
\end{eqnarray}
where constants $\dot{M}$ and $v_w$ are respectively the mass loss rate and wind velocity. We adopt a normalization parameter for the density $(\dot{M}/v_w)_*=10^{-4}M_\odot{\rm yr^{-1}}/({\rm km\ s^{-1}})$. The number density and velocity at the immediate downstream can be obtained from the self-similar solution of the shock dynamics \citep{Chevalier82}. The details of shock propagation are described in Appendix \ref{sec:dynamics}. We here overview how the dissipated kinetic energy at these shocks are converted to radiation and how the radiation emanates from the CSM. 

The immediate upstream, either the unshocked ejecta or unshocked CSM, is a cold but ionized gas due to strong UV/X-ray emission from the shocked region. When this crosses the shock, thermal relaxations between particles of the same species first occur; for a gas of density $\rho$ the electron-electron and proton-proton collision timescales \citep{Padmanabhan00}, 
\begin{eqnarray}
t_{\rm ee} &\approx& \frac{1}{4\pi}\frac{m_e^{1/2}(k_BT_e)^{3/2}}{e^4(\rho/m_p)\ln\Lambda} \nonumber\\
&\sim& 4\times 10^{-5}{\rm sec} \left(\frac{\rho}{10^{-13}\ {\rm g\ cm^{-3}}}\right)^{-1}\left(\frac{T_\mathrm{e}}{10^6{\rm K}}\right)^{3/2}  \label{eq:t_ee} \\
t_{\rm pp} &\sim&  \left(\frac{m_p}{m_e}\right)^{1/2}\left(\frac{T_p}{T_e}\right)^{3/2}t_{\rm ee} \nonumber \\
&\sim& 60\ {\rm sec} \left(\frac{\rho}{10^{-13}\ {\rm g\ cm^{-3}}}\right)^{-1}\left(\frac{T_\mathrm{p}}{10^9{\rm K}}\right)^{3/2} \label{eq:t_pp}
\end{eqnarray}
are generally much shorter than the dynamical timescales, 
\begin{equation}\label{eq:t_dyn}
    t_\mathrm{dyn} = \frac{r}{v_{\mathrm{fs}(rs)}} \sim 10\ {\rm days} \left(\frac{r}{10^{15}{\rm cm}}\right)\left(\frac{v_{\rm fs(rs)}}{10^{4}{\rm km\ s^{-1}}}\right)^{-1} .
\end{equation}
Here $k_{\rm B}$ is the Boltzmann constant, $m_e (m_p)$ is the electron (proton) mass, $e$ is the electron charge, and $\ln\Lambda\approx 30$ is the Coulomb logarithm. Then the temperature $T_\mathrm{e}$ ($T_\mathrm{p}$) of electrons (protons) at the immediate downstream of shock fronts is obtained as functions of the shock velocities $v_{\rm fs(rs)}$ from the jump condition with an adiabatic index $\gamma=5/3$ as
\begin{eqnarray}
    T_{\rm e,fs(rs)} &=& \frac{3}{16}\frac{m_e}{k_{\rm B}}v_{\rm fs(rs)}^2 \sim 10^6\ {\rm K}\left(\frac{v_{\rm fs(rs)}}{10^{4}{\rm km\ s^{-1}}}\right)^2 , \label{eq:Te_init}\\
    T_{\rm p,fs(rs)} &=& \frac{3}{16}\frac{ m_p}{k_{\rm B}}v_{\rm fs(rs)}^2 \sim 2\times 10^9\ {\rm K}\left(\frac{v_{\rm fs(rs)}}{10^{4}{\rm km\ s^{-1}}}\right)^2. \label{eq:Tp_init}
\end{eqnarray}
Most of the energy dissipated at the shock goes to protons at the immediate downstream.

The FS (RS) downstream is slightly slower (faster) than the contact discontinuity, so for both shocks the incoming gas is gradually advected to the contact discontinuity. 
The advection timescale is roughly equivalent to the dynamical timescale $t_{\rm ad} \approx t_\mathrm{dyn}$~\footnote{The width of the downstream is compressed by a factor $(\gamma+1)/(\gamma-1)$, but from the jump condition the downstream velocity in the shock rest frame is reduced by the same factor with respect to the upstream velocity.}.
During the advection, the electrons in the shocked region exchange their energies with the protons due to Coulomb interaction with an equipartition timescale of
\begin{eqnarray}\label{eq:t_ep}
    t_{\rm ep}&\sim& \left(\frac{m_p}{m_e}\right)^{1/2}t_{\rm pp} \nonumber \\
    &\sim& 2\times 10^3\ {\rm sec}\left(\frac{\rho}{10^{-13}\ {\rm g\ cm^{-3}}}\right)^{-1}\left(\frac{T_\mathrm{p}}{10^9{\rm K}}\right)^{3/2}  .
\end{eqnarray}
The electrons also either gain or lose energy by emitting or absorbing photons via bremsstrahlung processes and by inelastic Compton scattering~\citep{Weaver76,Katz10}.
Since the timescales in equations (\ref{eq:t_ee}) and (\ref{eq:t_pp}) are shorter than equation (\ref{eq:t_ep}) and other relevant cooling and heating timescales, the shocked regions can be treated as two temperature plasmas. 
On the other hand, since the advection of the shocked plasma (equation \ref{eq:t_dyn}) is the slowest process, we can approximate that the radiative shock structure is quasi-steady. 
For each epoch we obtain the hydrodynamic parameters at the shock interface (shock radius, velocity and upstream density), and use these to solve the electron/proton temperatures inside each region swept by the FS and RS.

Photons are predominantly generated by bremsstrahlung at the immediate shock downstream of the FS and RS. The spectral energy distributions are modified by Compton scattering and free-free and bound-free absorption while propagating through the shocked regions.  
Radiation escapes from both regions after a diffusion time passes, and half of the radiation that escapes the FS region reaches the observer (see Fig. \ref{fig:schematic}).
We solve the radiation transfer to calculate the escaping photon spectrum consistently with the two temperature structure in the shocked regions.

\subsection{Basic equations}
\label{sec:formulation}

In this section, we review the basic equations that govern the structure of the shocked region.

As the scale heights of the downstream parameters are much smaller than $r$, the geometric factor $r^2$ can be treated as a constant in the narrow shocked region. Assuming a steady state in the shock rest frame the hydrodynamic equations are described as (e.g. \citealt{Takei20})
\begin{eqnarray}
\frac{d}{dr}(\rho v) = 0, && \label{eq:number_equation} \\
\frac{d}{dr}(\rho v^2 + p) = 0, && \label{eq:momentum_equation} \\
\frac{d}{dr}\left[\rho v\left(\frac{1}{2}v^2+U+\frac{p}{\rho}\right)\right] &=& -\frac{d}{dr}\int d\nu\ F_{\rm \nu},
\label{eq:energy_equation}
\end{eqnarray}
where $\rho$ is the mass density of the shock downstream, and the pressure $p$ and specific internal energy $U$ are defined as
\begin{eqnarray}
p &=& \frac{\rho}{m_p}k_{\rm B}[xT_e+xT_p+(1-x)T_{\rm HI}] + \frac{1}{3}\int d\nu \ e_{\rm rad, \nu}, \\
\rho U &=& \frac{3}{2}\frac{\rho}{m_p}k_{\rm B}[xT_e+xT_p+(1-x)T_{\rm HI}] + \int d\nu \ e_{\rm rad, \nu}.
\end{eqnarray}
where $x$ is the ionization fraction. We set $x=1$ when protons and electrons have not reached equipartition yet. Once $T_e=T_p$, assuming local thermodynamic equilibrium the value of $x$ can be obtained from the Saha equation. The gas is assumed to consist purely of hydrogen for simplicity, with the temperature of neutral and ionized hydrogen are assumed to be equal, $T_{\rm HI}=T_p$, in the entire shocked region. $F_\nu$ and $e_{\mathrm{rad}, \nu}$ are the energy flux and density of radiation, respectively. Flux-limited diffusion (e.g. \citealt{Levermore81}) gives the flux as a function of $e_{\rm rad,\nu}$ by the following equation
\begin{equation}
    F_\nu = -\frac{c\lambda_\nu}{\kappa_\nu\rho}\frac{de_{\rm rad, \nu}}{dr}
    \label{eq:flux}
\end{equation}
where $\kappa_\nu$ is the opacity, $c$ is the speed of light, and $\lambda_\nu$ is a parameter between 0 and 1/3, the limits corresponding to the optically thin and thick case respectively.

When electrons and protons have different temperatures, the Coulomb interaction works between them to reduce the difference. The energy exchange rate of protons and electrons $P_{\rm e-p}$ [erg/cm$^3$/s] is \citep{Katz11}
\begin{eqnarray}
P_{\rm e-p} &\approx&  \ln \Lambda\sqrt{\frac{2}{\pi}}\frac{m_e}{m_p}k_{\rm B}(T_p-T_e)n_i^2\sigma_T c\left(\frac{k_{\rm B} T_e}{m_ec^2}\right)^{-3/2},
\label{eq:P_e_p}
\end{eqnarray}
where $n_i = x\rho/m_p$ is the number density of protons and electrons, and $\sigma_T$ is the Thomson cross section. 
Noting that electrons are dominantly affected by the photon field, the energy equation can be rewritten into the following equations for $T_e$ and $T_p$
\begin{eqnarray}
\frac{d}{dr}\left[\frac{5}{2}\frac{\rho k_{\rm B}}{m_p}vxT_e \right] &=& P_{\rm e-p} -\frac{d}{dr}\int d\nu\left[\frac{4}{3}ve_{\rm rad,\nu}+F_\nu \right],
\label{eq:dotTe_updated}\\
\frac{d}{dr}\left[\frac{5}{2}\frac{\rho k_{\rm B}}{m_p}vT_p \right] &=& -P_{\rm e-p}-\frac{d}{dr}\left(\frac{1}{2}\rho v^3\right).
\label{eq:dotTp_updated}
\end{eqnarray}

The equation of radiation transfer in spherical symmetry along the advection flow of the downstream plasma is
\begin{eqnarray}
\frac{1}{c}\frac{\partial I_\nu}{\partial t}+\cos\theta\frac{\partial I_\nu}{\partial r}+\frac{\sin\theta}{r}\frac{\partial I_\nu}{\partial\theta}=\frac{j_\nu}{4\pi}-\alpha'_{\nu} I_\nu,
\label{eq:rad_transfer_fundamental}
\end{eqnarray}
where $I_\nu$ is the specific intensity, $\theta$ is the angle between the light ray's propagation direction and the radial direction, and
\begin{eqnarray}
    j_{\rm \nu}&\equiv& j_{\rm \nu, ff}\\
    \alpha'_\nu &\equiv& \alpha_{\nu, \rm ff}-\frac{j_{\rm \nu, IC}}{ce_{\rm rad,\nu}},
\end{eqnarray}
are respectively the emissivity and absorption coefficient, which are related to the free-free and Comptonization processes. Taking the zeroth moment of equation (\ref{eq:rad_transfer_fundamental}) we obtain
\begin{eqnarray}
\frac{\partial e_{\rm rad,\nu}}{\partial t}+\frac{\partial F_\nu}{\partial r}=j_\nu-\alpha'_\nu ce_{\rm rad,\nu}.
\end{eqnarray}
Here we have neglected a geometric factor $2F_\nu/r$ because the relevant scale lengths are much shorter than the radius $r$.
Transferring to the Euler description and in the shock's rest frame, $\partial/\partial t \to (\partial/\partial t+v\partial/\partial r)$. From the steady state assumption we obtain
\begin{eqnarray}
v \frac{de_{\rm rad, \nu}}{dr} + \frac{dF_\nu}{dr} &=& j_{\rm \nu}-\alpha'_\nu ce_{\rm rad,\nu}.
\label{eq:rad_transfer}
\end{eqnarray}

For simplicity we assume free-free is the dominant emission/absorption processes and neglect free-bound/bound-free and bound-bound contributions. The emissivity [erg/s/cm$^3$/Hz] and absorption coefficient [cm$^{-1}$] are given in cgs units by \citep{Radipro}
\begin{eqnarray}
j_{\nu, \rm ff}&\approx& 6.8\times 10^{-38}n_i^2T_e^{-1/2}g_{\rm ff}\exp[-h\nu/k_{\rm B}T_e], \label{eq:j_ff}\\
\alpha_{\nu, \rm ff}&\approx&  3.7\times 10^{8}T_e^{-1/2}n_i^2\nu^{-3}g_{\rm ff}(1-\exp[-h\nu/k_{\rm B}T_e]),
\label{eq:alpha_ff}
\end{eqnarray}
where $h$ is Planck's constant and $g_{\rm ff}$ is the Gaunt factor
\begin{equation}
g_{\rm ff}(\nu)\sim {\rm max}\left(1, \frac{\sqrt{3}}{\pi}\ln\left[\frac{2.25k_{\rm B}T_e}{h\nu}\right]\right).
\end{equation}
For the photon frequency we sample a wide range of 0.1 eV$<h\nu<100$ MeV, with 80 bins evenly spaced on a logarithmic scale.

For inverse-Compton scattering, we obtain the rate of change in photon energy density $j_{\nu, \rm IC}$ [erg/s/cm$^3$/Hz] as 
\begin{eqnarray}
j_{\nu, \rm IC}&=& {\rm (gain\ term)\ - \ (loss\ term)} \nonumber \\
&\approx& \frac{\nu e_{\rm rad, \nu'}}{\nu'}[n_i\sigma_{\rm KN}(\nu')c] \frac{d\nu'}{d\nu}
- e_{\rm rad,\nu}n_i\sigma_{\rm KN}(\nu) c,
\end{eqnarray}
where $\sigma_{\rm KN}(\nu)$ is the Klein-Nishina cross section given by \citep{Radipro}
\begin{eqnarray}
    &&\frac{\sigma_{\rm KN}(\nu)}{\sigma_T}=\frac{3}{4}\times \nonumber \\
    &&\left[\frac{1+\beta}{\beta^3}\left\{\frac{2\beta(1+\beta)}{1+2\beta}-\ln(1+2\beta)\right\}+\frac{\ln(1+2\beta)}{2\beta}-\frac{1+3\beta}{(1+2\beta)^2}\right], \nonumber 
\end{eqnarray}
with $\beta=h\nu/m_ec^2$, and $\nu'(\nu,T_e)$ is the typical frequency of photons whose frequency becomes $\nu$ after scattering.

For a non-relativistic electron ($k_{\rm B}T_e\ll m_ec^2$) and a photon energy $h\nu'\ll m_ec^2$, the typical frequency $\nu$ after scattering of a photon of energy $\nu'$ is \citep{Radipro}
\begin{equation}
\nu = \nu' +  \frac{\nu'}{m_ec^2}(4k_{\rm B}T_e-h\nu').
\label{eq:nu_vs_nuprime_Radipro}
\end{equation}
Since we consider a photon spectrum at high energies exceeding $m_ec^2$ as well, we refine the second term as follows. For a photon scattering with an electron at rest, from the kinematics of the scattering
\begin{eqnarray}
\nu = \frac{\nu'}{1+(h\nu'/m_ec^2)(1-\cos\theta)}, \nonumber
\end{eqnarray}
where $\theta$ is the angle between the momenta of scattered and incident photons. Averaging over $\theta$
\begin{eqnarray}
\langle \nu \rangle &=& \frac{1}{2}\int_0^\pi \frac{\nu'\sin\theta d\theta}{1+(h\nu'/m_ec^2)(1-\cos\theta)} \nonumber \\
&=& \frac{m_ec^2}{2h}\ln\left(1+\frac{2h\nu'}{m_ec^2}\right).\nonumber
\end{eqnarray}
Repeating the same argument that led to equation (\ref{eq:nu_vs_nuprime_Radipro}), we obtain
\begin{equation}
    \nu = \frac{\nu'}{m_ec^2}\cdot 4k_{\rm B}T_e + \frac{m_ec^2}{2h}\ln\left(1+\frac{2h\nu'}{m_ec^2}\right).
    \label{eq:nu_of_nu'}
\end{equation}
The factor 4 is valid only when electrons are non-relativistic, which is generally satisfied for our model parameters. For each $\nu$ we solve equation (\ref{eq:nu_of_nu'}) implicitly and obtain $\nu'$.

\subsection{Two zone approximation}
\label{sec:twozone}
By solving equations (\ref{eq:number_equation}), (\ref{eq:momentum_equation}), (\ref{eq:flux}), (\ref{eq:dotTe_updated}), (\ref{eq:dotTp_updated}) and (\ref{eq:rad_transfer}) for both the forward and reverse shock regions, one can in principle obtain the shock structure, i.e., $T_\mathrm{e}(r)$, $T_\mathrm{p}(r)$, $\rho(r)$, $v(r)$, $F_\nu(r)$ and $e_{\rm rad, \nu}(r)$. These equations take into account the radiation feedback onto the dynamics of the advected gas, but this will be self-consistent only if the propagation of the shock is simultaneously modified \citep{Takei20}. Our formulation relies on the self-similar solution for the shock propagation and parameters at the immediate downstream, and taking into account the deviation from this self-similar evolution drastically complicates the problem.

In order to simplify the problem, in this work we adopt an assumption that the pressure gradient is zero in each shocked region, i.e. the dynamics is assumed to be one-zone. The plasma that crosses the shock is thus advected by a constant advection velocity, given by the downstream velocity in the shock's rest frame. \cite{Takei20}, which solved the above equations with the assumption of local thermodynamic equilibrium in the shocked region and adopting the diffusion approximation, shows that a nearly constant pressure is indeed seen (see their Figure 1). 

On the other hand we follow the evolution of $T_e, T_p$ and $e_{\rm rad,\nu}$ in the region, which are important for the emission. With our assumption the energy equation (\ref{eq:energy_equation}) is simplified to
\begin{equation}
    \frac{d}{dr}(\rho v U) = -\frac{d}{dr}\int d\nu F_\nu.
\end{equation}
Under the one-zone assumption, we average the flux that appears in the energy equation, over the shocked region's volume. The volume average then becomes a surface integral, and the energy equation is modified as
\begin{eqnarray}
\frac{d}{dr}\left(\rho v U\right) = \int d\nu (\dot{e}_{\rm esc,\nu}-j_{\nu, \rm in}),
\end{eqnarray}
where $e_{\rm esc,\nu}$ [${\rm erg/cm^3/Hz}$] is the energy density of radiation taken away from the shocked region, and $j_{\nu,\rm in}$ is the source term which consists of two components explained later in this section. 
From our constant pressure assumption $v$ is independent of $r$, and we can set $vd/dr=d/dt$, where $t$ is the time passed for the downstream plasma from when it crossed the shock. Then while $T_p>T_e$, the equations for $T_e$ and $T_p$ are
\begin{eqnarray}
\frac{3}{2}\frac{\rho k_{\rm B}}{m_p}\frac{d}{dt}(xT_e) &=& P_{\rm e-p} - \int d\nu \left(\dot{e}_{\rm rad,\nu}+\dot{e}_{\rm esc,\nu}-j_{\rm \nu, in}\right),
\\
\frac{3}{2}\frac{\rho k_{\rm B}}{m_p}\frac{{T}_p}{dt} &=&  -P_{\rm e-p}.
\end{eqnarray}
We make $T_e$ and $T_p$ equal by hand (with the total energy conserved) if these two approach within 10\%. This is merely for numerical purpose, as a finite difference tends to make the Coulomb term $P_{\rm e-p}$ erroneously large after electrons and protons have cooled down. 
Once $T_e=T_p$, thermal relaxation is fast enough that this equality is kept. Then the equations for $T_e$ and $T_p$ are further simplified to
\begin{eqnarray}
\frac{\rho k_{\rm B}}{m_p}\frac{d[(1+x)T_e]}{dt}= -\frac{2}{3}\int d\nu \left(\dot{e}_{\rm rad,\nu}+\dot{e}_{\rm esc,\nu}-j_{\rm \nu, in}\right).
\end{eqnarray}
We first obtain $[1+x(T_e)]T_e$ with this equation, then use the Saha equation and implicitly obtain $T_e(=T_p)$.

We next present our method of obtaining $e_{\rm rad,\nu}, e_{\rm esc,\nu}$. The equation for radiation transfer becomes
\begin{eqnarray}
\dot{e}_{\rm rad, \nu} + \dot{e}_{\rm esc, \nu}  &=& (j_{\nu, \rm ff}+j_{\nu, \rm in})-\alpha'_\nu ce_{\rm rad,\nu}, \nonumber \\
&\equiv& j_{\nu}-\alpha'_\nu ce_{\rm rad,\nu},
\label{eq:rad_transfer_time}
\end{eqnarray}
where we redefined $j_{\nu}\equiv j_{\nu, \rm ff}+j_{\nu, \rm in}$. This equation reduces to the standard equation of radiation transfer if $\dot{e}_{\rm esc,\nu}=0$ is assumed. In this case the formal solution can be obtained by expressing $e_{\rm rad,\nu}$ as $e_{\rm rad,\nu}(t)=F(t)\exp[-\int^t_0 ds \ \alpha'_{\nu}(s)c]$, and solving for $F(t)$ by substituting this into equation (\ref{eq:rad_transfer_time}).

To determine $\dot{e}_{\rm rad,\nu}$ and $\dot{e}_{\rm esc,\nu}$ separately we need another constraint, which should be related to the physics of diffusion. As an approximation we treat the diffusion as a hard cutoff over a typical diffusion timescale $t_{\rm diff}$. In other words, photons created at time $t=s$ are in the shocked region and contribute to $e_{\rm rad,\nu}$ at time $s<t<s+t_{\rm diff}$, then completely escape and contribute to $e_{\rm esc,\nu}$ at $t>s+t_{\rm diff}$. 
Then by analogy of the standard case described above, we find the solution of the form
\begin{eqnarray}
e_{\rm rad, \nu}(t)&=&\int_{t-t_{\rm diff}}^{t} j_{\nu}(s)ds\exp\left[-\int_s^t\alpha'_\nu(\tilde{s})c d\tilde{s}\right],
\label{eq:erad}\\
e_{\rm esc, \nu}(t)&=& \int_0^{t-t_{\rm diff}} j_{\nu}(s)ds \exp\left[-\int_s^{s+t_{\rm diff}} \alpha'_\nu(\tilde{s})c d\tilde{s}\right],
\label{eq:eesc}
\end{eqnarray}
for $t>t_{\rm diff}$, and
\begin{eqnarray}
e_{\rm rad, \nu}(t)&=&\int_{0}^{t} j_{\nu}(s)ds\exp\left[-\int_s^t\alpha'_\nu(\tilde{s})c d\tilde{s}\right], \label{eq:erad_bef_tdiff} \\
e_{\rm esc, \nu}(t)&=& 0,
\end{eqnarray}
for $0<t\leq t_{\rm diff}$. To calculate the integrals, we keep records of $j_{\rm \nu}$ and $\alpha'_{\rm \nu}$ throughout the dynamical time with timestep $t_{\rm dyn}/24000$. This resolution is limited by our computational resource. In the RS region, we observed cases where the cooling timescale (equation \ref{eq:t_cool_rs}) became shorter than this time resolution, thereby preventing us to obtain these integrals accurately. In these cases we switched to an analytical treatment for calculating $e_{\rm rad,\nu}$ and $e_{\rm esc,\nu}$, explained in Appendix \ref{sec:Ultra-fast cooling}.

We calculate $T_e, T_p, e_{\rm rad,\nu},$ and $e_{\rm esc,\nu}$ from $t=0$ to $t_{\rm end}$, the timescale for the gas to be advected from the shock to the contact discontinuity, in our case equivalent to the dynamical time (see Section \ref{sec:summary}). Thus $t_{\rm end}$ is roughly equivalent to the epoch $\mathcal{T}$, or time from explosion\footnote{This is not to be confused with the variable $t$, which was previously defined for the calculation of the shock structure at a given epoch $\mathcal{T}$.}. We do this integration for a series of epochs  $\{\mathcal{T}_0,\mathcal{T}_1,\cdots\}$, with $t_{\rm end,N}=\mathcal{T}_N\ (N=0,1,\cdots)$. For each epoch the conditions at $t=0$ (i.e. at the shock) are obtained by the equations in Appendix \ref{sec:dynamics}. 
The series should be set so that the dynamics and radiation flow do not significantly change between neighboring epochs. We set the epochs by the following equations
\begin{eqnarray}
\mathcal{T}_0&=&{\rm max}(1\ {\rm day},1.1\mathcal{T}_{\rm CSM})\nonumber \\
\mathcal{T}_{N+1}-\mathcal{T}_{N}
&=& \left\{ \begin{array}{ll}
0.5t^{\rm FS}_{\rm diff, N} & (\mathcal{T}_N<10t^{\rm FS}_{\rm diff, N}),\\
0.1\mathcal{T}_N  & (\mathcal{T}_N\geq 10t^{\rm FS}_{\rm diff, N}),
\end{array}\right.
\end{eqnarray}
The parameter $\mathcal{T}_{\rm CSM}$ is the diffusion timescale in the CSM which controls the light curve evolution
\begin{eqnarray}
    \mathcal{T}_{\rm CSM} &\approx& \frac{\kappa_{\rm scat} \dot{M}}{v_w c} \nonumber \\
    &\sim& 0.7\ {\rm day}\left(\frac{\kappa_{\rm scat}}{0.34\ {\rm cm^{2} g^{-1}}}\right)\left[\frac{\dot{M}/v_w}{(\dot{M}/v_w)_*}\right]
\end{eqnarray}
where $\kappa_{\rm scat}\approx 0.34\ {\rm cm^2\ g^{-1}}$ is the Thomson scattering opacity. The series continues up to 250 days or when the RS sweeps the entire outer ejecta, whichever is earlier. Variables $t^{\rm RS}_{\rm diff}, t^{\rm FS}_{\rm diff}$ are respectively the diffusion timescales in the RS and FS region. The diffusion time in the FS region $t^{\rm FS}_{\rm diff}$ is given by its optical depth multiplied by the light-crossing timescale
\begin{equation}
t^{\rm FS}_{\rm diff}\sim {\rm min}[\kappa_{\rm scat}\rho_{\rm fs}\Delta r_{\rm FS}, 1](\Delta r_{\rm FS}/c),
\end{equation}
where the width of the FS region $\Delta r_{\rm FS}$ relates to the radius of the  contact discontinuity  $r_{\rm sh}$ derived in equation (\ref{eq:rsh_ss}) as $\Delta r_{\rm FS}\approx [(\gamma-1)/(\gamma+1)]r_{\rm sh}$. The diffusion time in the RS region $t^{\rm RS}_{\rm diff}$ depends on the parameters for the self-similar solution, but is generally much shorter than $t^{\rm FS}_{\rm diff}$ due to the much smaller width. For example if we assume $n=10$ and $\gamma=4/3$, we find $t^{\rm RS}_{\rm diff}\approx0.134t^{\rm FS}_{\rm diff}$.

In our model $j_{\nu, \rm in}$ has two components. One is radiation entering from the other shocked region, shown as black arrows in Figure \ref{fig:schematic}. For radiation escaping from the FS region, assuming the escaping radiation is isotropic $r_{\rm fs}^2/(r_{\rm sh}^2+r_{\rm fs}^2)\approx 1/2$ escapes outwards (which is observed), while the other 1/2 goes inwards to the RS region, and is possibly reprocessed there. Under the assumption of cold SN ejecta in which only the vicinity of the shocked region is ionized, we assume that the bulk of the ejecta are transparent and all the escaping radiation from the RS region goes to the FS region as a source term. We calculate the source term from the $e_{\rm esc,\nu}$ in the previous epoch, and inject this at a constant rate for $t_{\rm dyn}$.

The other component, which is rather artificial, is $e_{\rm rad,\nu}$ that should be present in the shocked region at $t=0$. Our formulation of $e_{\rm rad,\nu}$ (equation \ref{eq:erad_bef_tdiff}) implicitly assumes $e_{\rm rad,\nu}=0$ at $t=0$, which neglects this contribution. We instead include this as a source term estimated from the previous epoch; i.e. the leftover $e_{\rm rad,\nu}$ in the end of the previous epoch, $e_{\rm rad,\nu}(t=t_{\rm end, N-1})$, is carried over in the present epoch $\mathcal{T}_N$ as a source term. We inject this at constant rate for the first diffusion time at epoch $\mathcal{T}_N$.

When setting the injection term from the previous epoch, we make sure that the total injected energy in the shocked region, $e_{\rm esc,\nu}$ (or $e_{\rm rad,\nu}$) times the volume of the shocked region, is conserved. Mathematically, $j_{\rm \nu, in}$ is given for the FS and RS regions as
\begin{widetext}
\begin{equation}
j_{\nu,\rm in}^{\rm FS} = \frac{V^{\rm FS}_{\rm N-1}}{V^{\rm FS}_{\rm N}} \left[\left(\frac{V^{\rm RS}_{\rm N-1}}{V^{\rm FS}_{\rm N-1}}\right)\frac{e_{\rm esc,\nu}^{\rm RS}(t=t_{\rm end, N-1})}{\mathcal{T}_{N}}+\frac{e^{FS}_{\rm rad,\nu}(t=t_{\rm end, N-1})}{t^{FS}_{\rm diff, N}}\Theta(t^{FS}_{\rm diff,N}-t)\right], 
\end{equation}
\begin{equation}
j_{\nu, \rm in}^{\rm RS} = \frac{V^{\rm RS}_{\rm N-1}}{V^{\rm RS}_{\rm N}} \left[\left(\frac{V^{\rm FS}_{\rm N-1}}{V^{\rm RS}_{\rm N-1}}\right)\frac{e_{\rm esc,\nu}^{\rm FS}(t=t_{\rm end, N-1})}{2\mathcal{T}_{N}}+\frac{e^{RS}_{\rm rad,\nu}(t=t_{\rm end, N-1})}{t^{RS}_{\rm diff, N}}\Theta(t^{RS}_{\rm diff,N}-t)\right],
\end{equation}
\end{widetext}
where $\Theta(t)$ is the Heaviside step function, and $V^{\rm FS(RS)}_N$ is the volume of the FS (RS) region at epoch $\mathcal{T}_N$. Variables with superscript FS (RS) denote values in the FS (RS) region. In the self-similar solution the ratio of the volumes of the FS and RS regions is a constant of time.
We neglect additional energy sources that can appear as a source term, such as radioactive decay of nickel inside the ejecta.

For the first epoch $\mathcal{T}_0$, the calculation is iterated until $e_{\rm rad,\nu}$ becomes stable. To obtain $j_{\rm \nu, in}$ for the next iteration, we use the values of $e_{\rm rad,\nu}$ and $e_{\rm esc, \nu}$ at $t=t_{\rm end,0}$ because there is no previous epoch. The energy density at $t=t_{\rm end,0}$ becomes stable within 1\% after about 10 iterations, from which we move on to epoch $\mathcal{T}_1$.

\subsection{Calculating the Observed Spectrum}
To obtain the observed spectrum, we need to solve the radiation transfer through the unshocked CSM.
At each epoch, the luminosity of radiation that escapes from the FS region towards the observer is
\begin{eqnarray}
L_\nu(\mathcal{T}_N) &\approx& \frac{1}{2}V_{\rm N}^{\rm FS}\frac{e^{\rm FS}_{\rm esc, \nu}(t=t_{\rm end, N})}{\mathcal{T}_N} \nonumber \\ \nonumber \\
&\approx& 2\pi r_{\rm sh}^2\Delta r_{\rm FS} \frac{e^{\rm FS}_{\rm esc, \nu}(t=t_{\rm end,N})}{\mathcal{T}_N},
\end{eqnarray}
where the factor 1/2 is due to the aforementioned geometrical effect. Once we obtain this luminosity $L_{\nu}$, we use version 17.02 of CLOUDY \citep{Ferland17} to obtain the spectrum after being processed through the unshocked CSM $L_{\rm \nu, CSM}$. We assume the CSM extends with a spherically symmetric wind profile from $r=r_{\rm sh}$ out to $10^{17}$ cm. The metallicity of the CSM is assumed to be solar.

This treatment also assumes a steady state, i.e. the photoionization at a given epoch is determined by the luminosity at that time. For Type IIn SNe this is justified, as the recombination timescale is generally much shorter than the dynamical timescale \citep{Chevalier12}. For example, for a gas of $T_e \sim 10^4$ K the recombination coefficient is $\alpha_{\rm rec}\sim O(10^{-13})\ {\rm cm^{3}\ s^{-1}}$, and the recombination timescale is of order $10^3$ ($10^5$) s at day 10 (100) for our fiducial model parameters.

We also include the effect of diffusion inside the CSM delaying the light curve, by the following formula (\citealt{Chatzopoulos12}; equation 4)
\begin{eqnarray}
    L_{\rm \nu, obs}(\mathcal{T})=\frac{e^{-\mathcal{T}/\mathcal{T}_{\rm CSM}}}{\mathcal{T}_{\rm CSM}}\int_0^\mathcal{T} d\mathcal{T}'e^{\mathcal{T}'/\mathcal{T}_{\rm CSM}}L_{\rm \nu, CSM}(\mathcal{T}')
\end{eqnarray}
For the contribution of $L_{\nu, \rm CSM}$ at $t<\mathcal{T}_0$, which is absent in our modelling, we adopt the value at $t=\mathcal{T}_0$ for simplicity. The actual contribution is uncertain, as this also depends on when the CSM interaction starts. However, the exponential dependence makes this uncertainty unimportant after about a few times $\mathcal{T}_{\rm CSM}$.

\section{Results}
\label{sec:results}

\begin{table*}
\centering
\begin{tabular}{c|ccc}
Model & $E_{\rm ej}$ [erg] & $\dot{M}/v_w\ [(M_\odot/{\rm yr})/({\rm km/s})]$ & $M_{\rm CSM}(r<10^{16}{\rm cm})$ [$M_\odot$]  \\ \hline
E1M1 & $10^{51}$& $1\times 10^{-2}/100$ & 0.3 \\
E3M1 & $3\times 10^{51}$& $1\times 10^{-2}/100$ & 0.3 \\
E10M1 & $10^{52}$& $1\times 10^{-2}/100$ & 0.3\\
E1M10 & $10^{51}$& $1\times 10^{-1}/100$ & 3 \\
E3M10 & $3\times 10^{51}$& $1\times 10^{-1}/100$ & 3 \\
E10M10 & $10^{52}$& $1\times 10^{-1}/100$ & 3\\
\end{tabular}
\caption{Summary of model parameters adopted in this work. The parameters are explosion energy $E_{\rm ej}$ and wind density parameter $\dot{M}/v_w$. The last column shows the enclosed mass of CSM within $10^{16}$ cm.} 
\label{table:Parameters}
\end{table*}

We consider the model parameters listed in Table \ref{table:Parameters}, which are within the range of typical inferred parameters for Type IIn SNe \citep{Smith14}. Model E$\epsilon$M$\mu$ represents a model with ejecta energy $\epsilon\times 10^{51}$ erg, and CSM density parameter $\dot{M}/v_w=\mu(\dot{M}/v_w)_*$. We assume $n=10$ and adiabatic index $\gamma=4/3$ for all of our parameter sets when solving the shock dynamics from the self-similar solution. We have assumed the ejecta mass to be $10\ M_\odot$ for all cases.

We first discuss the shock downstream, such as the evolution of the electron temperature and contribution of various heating/cooling process. These processes are also important for shaping the radiation spectrum. We then discuss the spectra and the optical and X-ray light curves obtained from our calculations.

\subsection{Shock Downstream}

\begin{figure*}[t]
\centering
\includegraphics[width=\linewidth]{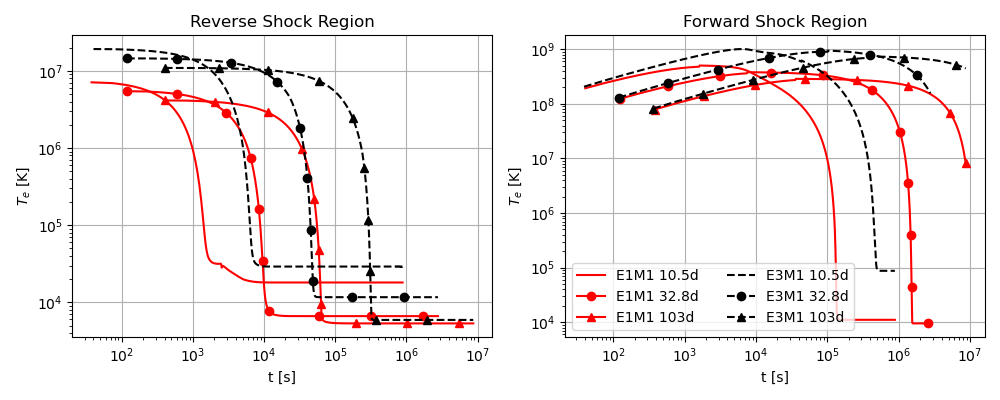}
\caption{Time evolution of the electron temperature in the RS (left panel) and FS (right panel) regions, for the E1M1 (solid) and E3M1 (dashed) models and three epochs.}
\label{fig:Te_evolution}
\end{figure*}

\begin{figure*}[t]
\centering
\includegraphics[width=0.9\linewidth]{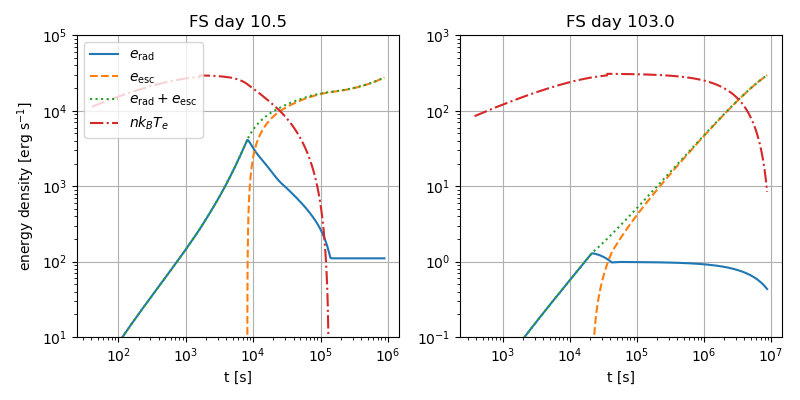}
\centering
\includegraphics[width=0.9\linewidth]{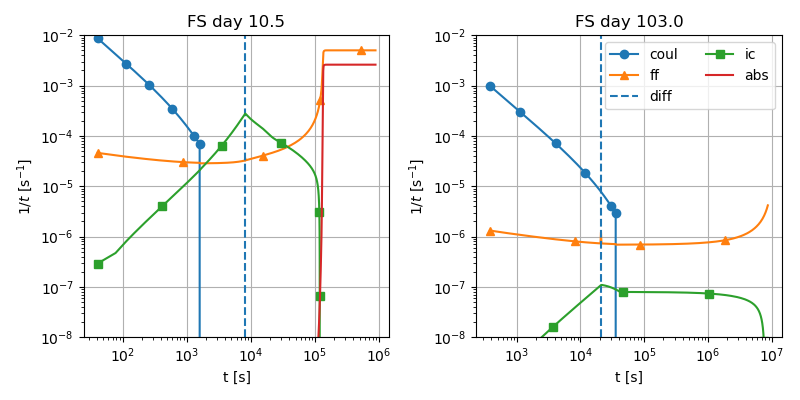}
\caption{(Top panel) Energy densities of radiation and electrons. The solid and dashed lines are $e_{\rm rad}$ and $e_{\rm esc}$, and the dotted line is their sum. (Bottom panel) Inverse of various cooling and heating timescales at the FS region for the E1M1 model. At each timestep the highest curve is the most effective heating/cooling process for changing the electron temperature. The vertical dashed line is the diffusion time at that epoch.}
\label{fig:FS_timescales}
\end{figure*}

We show the change of electron temperature $T_e$ as a function of time in Figure \ref{fig:Te_evolution}. The value of $T_e$ initially rises due to Coulomb interaction with the protons (hotter by $m_p/m_e$), but then the temperature drops first due to free-free emission, then by inverse Compton scattering. The detailed evolution differs for RS and FS, which we discuss in detail below.

\subsubsection{Reverse shock region}
In the RS region bremsstrahlung tends to be the dominant process for cooling electrons, owing to the high density and low shock velocity \citep{Chevalier12}. The available energy density at shock downstream is
\begin{eqnarray}
\frac{3}{2}n_ik_{\rm B}T_{p, \rm rs}&\approx& \frac{9}{14\alpha}\frac{\dot{M}}{4\pi v_w}\left(\frac{v_{\rm rs}}{r_{\rm sh}}\right)^2 \nonumber \\
&\sim& 10^3{\rm erg\ cm^{-3}}\left(\frac{\mathcal{T}}{10\ {\rm day}}\right)^{-2}\left[\frac{\dot{M}/v_w}{(\dot{M}/v_w)_*}\right]
\end{eqnarray}
where $\alpha$ is a parameter that appears in the self-similar solution of shock dynamics, as can be seen in equation (\ref{eq:rsh_ss}).
This thermal energy density corresponds to $5\times 10^6$ K for the E1M1 model at 10 days. The cooling timescale is dominant at the immediate downstream when $T_e$ is highest, and scales with time and model parameters as
\begin{eqnarray}
    t_{\rm cool} &\sim& \frac{(3/2)n_ik_{\rm B}T_{p,\rm rs}}{1.4\times 10^{-27}{\rm cgs}\ \bar{g}_{\rm ff}T_{p, \rm rs}^{1/2}n_i^2} \nonumber \\
    &\sim& 300\ {\rm s}\ \bar{g}_{\rm ff}^{-1} \left(\frac{\mathcal{T}}{10\ {\rm day}}\right)^{13/8}\nonumber \\
&&\left(\frac{E_{\rm ej}}{10^{51}{\rm erg}}\right)^{21/16} \left[\frac{\dot{M}/v_w}{(\dot{M}/v_w)_*}\right]^{-11/8},
\label{eq:t_cool_rs}
\end{eqnarray}
where $\bar{g}_{\rm ff}\approx 1.2$ is the frequency-averaged value of $g_{\rm ff}$ \citep{Radipro}. This timescale is consistent with the curves in the left panel of Figure \ref{fig:Te_evolution}. This $t_{\rm cool}$ is generally much smaller than the dynamical time $t$, and the plasma efficiently cools down and is eventually balanced with heating by free-free absorption. In this case the gas energy density is converted to black-body radiation whose temperature is given as
\begin{eqnarray}
T_{\rm BB} &\sim& \left[\frac{(3/2)n_ik_{\rm B}T_{p, \rm rs}}{a}\right]^{1/4} \nonumber \\
&\sim& 2\times 10^4\ {\rm K}\left(\frac{\mathcal{T}}{10\ {\rm day}}\right)^{-1/2} \left[\frac{\dot{M}/v_w}{(\dot{M}/v_w)_*}\right]^{1/4}.
\end{eqnarray}
where $a$ is the radiation density constant. There is also a contribution from the FS region, which can further increase $T_e$ due to absorption and interaction of FS photons with the electrons. The Compton heating is more important for models with higher ejecta energy, as it makes the seed photons harder. For some cases this balances free-free cooling and halts the decrease of $T_e$. 

The escape of photons after a diffusion time contributes to lowering $T_e$, as the supply of hard photons emitted in the RS region or streaming from the FS region would be limited. This generally makes $T_e<T_{\rm BB}$, and for some cases even reaching to temperatures of around $6000$ K where hydrogen recombines.

\subsubsection{Forward shock region}
The situation can be different for the FS region, due to the larger downstream energy density and, more importantly, the much lower density at the shock front. The free-free cooling time is orders of magnitude longer than in the RS region, with a scaling
\begin{eqnarray}
    t_{\rm cool} 
    &\sim& 4\times 10^4\ {\rm s}\ \bar{g}_{\rm ff}^{-1} \left(\frac{\mathcal{T}}{10\ {\rm day}}\right)^{13/8}\nonumber \\
&&\left(\frac{E_{\rm ej}}{10^{51}{\rm erg}}\right)^{21/16} \left[\frac{\dot{M}/v_w}{(\dot{M}/v_w)_*}\right]^{-11/8}.
\end{eqnarray}
This roughly explains the drop at later epochs, such as day 103. However at earlier epochs, due to the inefficient free-free cooling inverse Compton can take over the cooling when there are enough supply of seed photons \citep{Chevalier12}.

To see this more quantitatively, in the bottom panel of Figure \ref{fig:FS_timescales} we plot the time evolution of the inverse of various electron heating and cooling timescales in the E1M1 model. The Coulomb heating timescale is $t_{\rm e-p} \sim (3n_ik_{\rm B}T_e/2)/P_{\rm e-p}$, IC cooling timescale is $t_{\rm IC}\sim (3n_ik_{\rm B}T_e/2)/[\int d\nu j_{\nu, \rm IC}]$, and the free-free absorption heating timescale is $t_{\rm abs}\sim (3n_ik_{\rm B}T_e/2)/[\int d\nu\alpha_{\nu,\rm ff} c e_{\rm rad,\nu}]$. The aforementioned transition of IC cooling being the dominant cooling process can be seen in the bottom panels, at $\sim10^4$s in day 10.5. For early phases, this is likely to be  responsible for significantly reducing the electron temperature (see top panel of Figure \ref{fig:FS_timescales}) before free-free cooling takes back the role.

In the early epochs we see convergence of $T_e$, which is due to either (i) balance of free-free absorption and cooling given that the density is high enough, (ii) balance of free-free cooling and Compton heating by photons from the RS, or (iii) recombination of hydrogen for those that converge around $6000$ K. 

\subsection{Spectral Properties}

\begin{figure*}[t]
 \begin{tabular}{cc}
\begin{minipage}{0.5\hsize}
 \centering
\includegraphics[width=\linewidth]{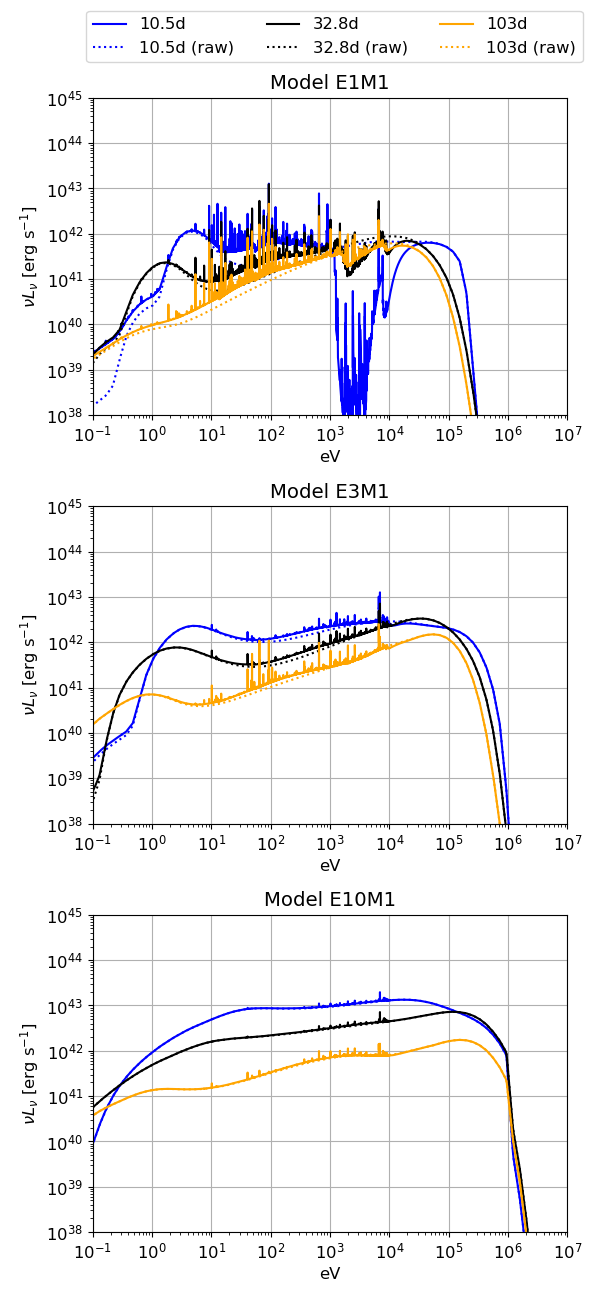} 
\end{minipage}

\begin{minipage}{0.5\hsize}
 \centering
\includegraphics[width=\linewidth]{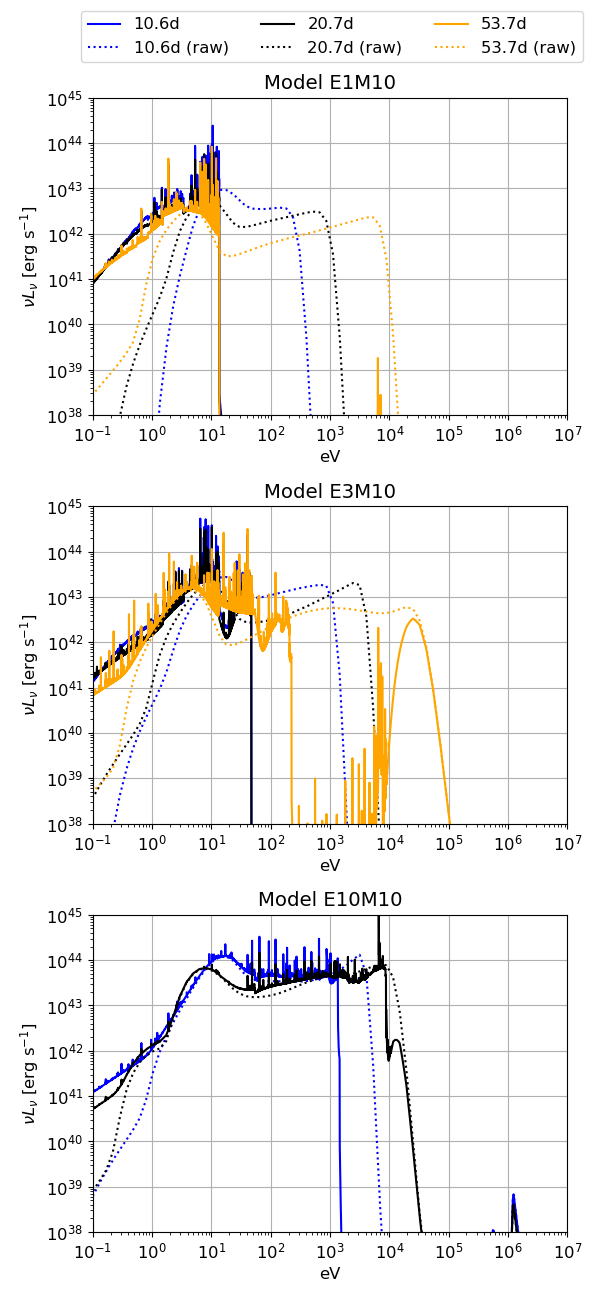}
\end{minipage}
\end{tabular}
\caption{Spectrum $\nu L_\nu$ for the models. The left 3 panels are for 'M1' models (from top 'E1', 'E3', 'E10'), and right panels for 'M10' models. The solid (dotted) line shows the spectrum after (before) being processed with CLOUDY.}
\label{fig:spectrum}
\end{figure*}

Figure \ref{fig:spectrum} shows the resulting spectra of photons escaping the FS region for our choice of model parameters. We find that for the E1M1 and all the M10 models soft X-rays emitted from the shock are significantly absorbed by the unshocked CSM, and the observed spectrum is much different from the spectrum emergent from the shock front. A fraction of the absorbed X-ray photons are converted to optical and UV emission, with a large number of emission lines including H$\alpha$ at $\approx 1.9$ eV. This matches with the naive expectation that for low $E_{\rm ej}$ and/or high $\dot{M}/v_w$, the shock expands slower and the optical depth of the CSM is kept high. On the other hand, the observed spectra for models E3M1 and E10M1 are largely unmodified from the spectrum from the shock. In addition to the CSM quickly becoming optically thin, larger explosion energies (hence larger shock velocity) extend the FS spectra to hard X-rays, which avoid significant absorption by the CSM.

An important aspect that determines the spectrum from the shock is the optical depth to Thomson scattering inside the FS region. For a wind profile CSM this is equivalent to that in the unshocked CSM, and is given as
\begin{eqnarray}
    \tau_{\rm fs}&\sim& \kappa_{\rm scat}\rho_w r_{\rm sh} \sim 1.7\left[\frac{\dot{M}/v_w}{(\dot{M}/v_w)_*}\right]\left(\frac{r_{\rm sh}}{10^{15}{\rm cm}}\right)^{-1},
\end{eqnarray}
where we assumed $\kappa_{\rm scat}=0.34\ {\rm cm^{2}g^{-1}}$. 

For an optically thick ($\tau_{\rm fs}\gg 1$) case, Compton downscattering can drastically reduce the photon energy. The number of Thomson scatterings per unit time is $\kappa_{\rm scat}\times 7\rho_{\rm w}c$, and photons stay in the FS region for a diffusion time $t_{\rm diff, fs}\sim \kappa_{\rm scat}\rho_{\rm w} r_{\rm sh}^2/7c$, where the factor 7 comes from the compression ratio for adiabatic index $4/3$. Thus there are $\tau_{\rm fs}^2$ scatterings in the FS region before escaping, and as a result the energy of photons reaching the observer is reduced to $\approx 511{\rm keV}/\tau_{\rm fs}^2\sim 0.3\ {\rm keV}(\tau_{\rm fs}/40)^{-2}$. This is consistent with the cutoff in the spectra at the FS for the E1M10 model on day 10.6, when $\tau_{\rm fs}\approx 38$ (blue dotted line in the top-right panel of Figure \ref{fig:spectrum}). 

The degree of absorption of soft X-rays depends on this cutoff frequency and the luminosity of ionizing photons $L_{\rm ion}$, as discussed in \cite{Chevalier12}. Qualitatively, in order for soft X-rays to reach the observer the following two conditions must be satisfied; (i) $\tau_{\rm fs}$ should be at most a few tens, so that the cutoff is in or beyond the soft X-ray band (ii) the flux of ionizing photons is high enough that the main elements that contribute to absorption of soft X-rays (oxygen and iron) are completely ionized. The latter is quantified as the ionization parameter $\xi\equiv L_{\rm ion}/(nr^2)\approx 4\pi m_pL_{\rm ion}/(\dot{M}/v_w)$ in cgs units. \cite{Chevalier12} finds that $\xi$ greater than $5000$--$10^4$ will ionize all the elements, a precise value depending on the characteristic temperature of the spectrum. For our parameters
\begin{eqnarray}
    \xi \sim  3\times 10^3 \left(\frac{L_{\rm ion}}{10^{43}\ {\rm erg\ s^{-1}}}\right) \left[\frac{\dot{M}/v_w}{(\dot{M}/v_w)_*}\right]^{-1},
\end{eqnarray}
which is consistent with our results within an order of magnitude, and explains the tendency seen in our spectra that models with larger $E_{\rm ej}$ (i.e. larger $L_{\rm ion}$) and smaller $\dot{M}/v_w$ are visible in soft X-rays.

Another notable feature is that the spectrum $\nu L_\nu$ below the cutoff is nearly flat for the E1M1 and E3M1 cases, while it becomes steeper up to $\nu L_\nu\propto \nu$ for the E10M1 case. The latter is simply from the fact that the free-free emissivity $j_{\nu,\rm ff}$ is flat at $h\nu\ll k_{\rm B}T_e$, and Compton and IC scatterings are ineffective. Repeated scattering between electrons and photons can convert X-ray photons to lower energies, which flattens the spectrum from the free-free one. 

\subsection{Optical and X-ray Light Curves}
\begin{figure*}
 \centering
\includegraphics[width=\linewidth]{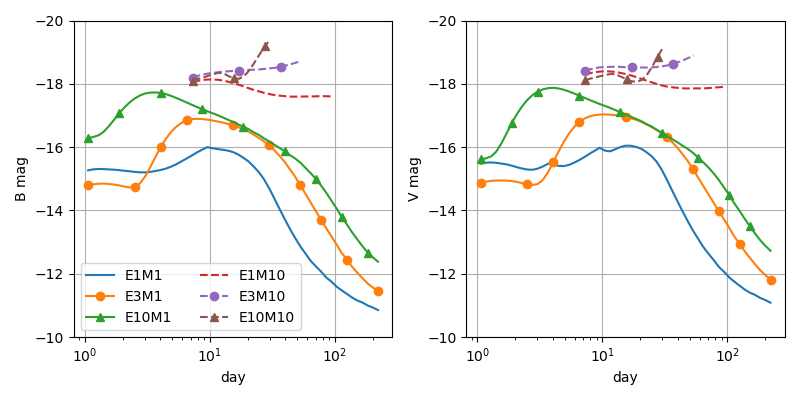}
\caption{B (left) and V (right) band light curves.}
\label{fig:B_V_band}
\end{figure*}

We calculate the optical and X-ray light curves to compare with observations. For the optical light curves, we use the $\nu L_\nu$ from our model and adopt the Johnson-Cousins filter response function to obtain the (absolute) B and V band magnitudes with
\begin{equation}
    M = -2.5\log_{10}\left\{\frac{\int [L_\nu/4\pi (10{\rm pc})^2] (c/\lambda^2)(\lambda/hc) e(\lambda)d\lambda }{\int f_{\nu, \rm ref}(c/\lambda^2)(\lambda/hc)e(\lambda)d\lambda}\right\},
\end{equation}
where $e(\lambda)$ is the filter response function for each band\footnote{https://www.aip.de/en/research/facilities/stella/instruments/data}, $f_{\nu, \rm ref}$ is the reference flux that defines the zero point, and we used the conversion $f_{\lambda}=(c/\lambda^2)f_\nu$. For $f_{\nu, \rm ref}$ we adopt the values $4260$ Jy and $3640$ Jy for the B and V bands respectively (see Table IV of \citealt{Bessell79}). 
The resulting light curves in B and V bands are shown in Figure \ref{fig:B_V_band}. The peak magnitude of average Type IIn SNe is estimated to be $M_V=-18.4\pm 1.0$ \citep{Kiewe12}, with more recent estimates \citep{Richardson14,Nyholm20} giving consistent results. Our results that roughly reproduced this magnitude favors mass-loss rates of order $0.1M_\odot\ {\rm yr}^{-1}$. This is at odds with the much lower $\dot{M}$ derived for some SNe IIn in previous works \citep{Kiewe12,Taddia13}. This should be partly due to the shortcomings of our model with incomplete opacity in the shocked region and an assumption of a one-zone density profile, which are elaborated in Section \ref{sec:caveats}. The discrepancy can also be from their assumption of a high conversion efficiency from kinetic energy dissipation rate to H$\alpha$ luminosity, $\epsilon_{\rm H\alpha}=0.1$. Inspecting Figure \ref{fig:spectrum}, some spectra show $\epsilon_{\rm H\alpha}\ll 0.1$, especially for the M1 models. A low $\epsilon_{\rm H\alpha}$ is also implied from a more recent SN IIn sample \citep{Kokubo19}. As the value of $\dot{M}$ derived from this method scales as the inverse of $\epsilon_{\rm H\alpha}$, this implies that the actual $\dot{M}$ can be much higher than previously estimated. A more sophisticated modelling of $\epsilon_{\rm H\alpha}$ may give a better understanding on $\dot{M}$, although it is beyond the scope of this work.

\begin{figure*}
 \centering
\includegraphics[width=\linewidth]{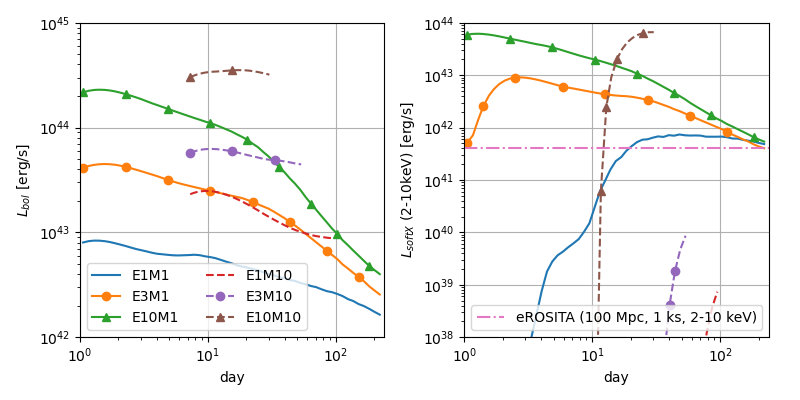}
\caption{Bolometric and X-ray light curves. Overlaid as dash-dot line is the sensitivity of eROSITA for a source at distance 100 Mpc, with $10^3$ sec exposure time in energy range 2-10 keV (\citealt{Merloni12}; Figure 4.3.1).}
\label{fig:lightcurve}
\end{figure*}

For the X-ray luminosity, most of the observations of SNe IIn, using e.g. Chandra and XMM-Newton, have been done in the soft X-ray band covering energies up to $10$ keV. Here we define the X-ray luminosity as
\begin{eqnarray}
L_{\rm X} = \int_{2{\rm keV}}^{10{\rm keV}}L_\nu d\nu.
\end{eqnarray}
This is to be compared with the bolometric luminosity integrated over the frequency range in our model
\begin{eqnarray}
L_{\rm bol} = \int_{0.1{\rm eV}}^{10^7 {\rm eV}}L_\nu d\nu.
\end{eqnarray}
We show the resulting light curves in Figure \ref{fig:lightcurve}. For the E1M1, E1M10 and E3M10 models, CSM absorption is significant enough to make X-rays observable only at later times. The E1M10 model is extreme, as soft X-ray photons are completely absorbed until the end of our simulation of $\sim 200$ days. For most SNe IIn X-rays have been observed only around a year after explosion \citep{Chandra18}. This observational fact may favor mass-loss rates of order $0.1M_\odot\ {\rm yr}^{-1}$ for SNe IIn, independent from the previous argument on the optical luminosity. For models E3M1 and E10M1, where the CSM quickly becomes transparent to X-rays, the X-ray luminosity can become comparable to the bolometric luminosity. In fact, they show very bright X-ray emission of luminosity $10^{42}$--$10^{44}\ {\rm erg\ s^{-1}}$ for $\sim 100$ days. While the rate of such energetic explosions inside a moderate CSM is uncertain, they are good targets for all-sky X-ray surveys carried out with e.g. eROSITA (see Figure 5.7.2 of \citealt{Merloni12}).

\section{Comparison with Type IIn Supernovae}
\label{sec:comparison}

As a demonstration of our emission model, we test our model with observations of two well-studied Type IIn SNe, SN 2010jl and SN 2014C.

\subsection{SN 2010jl}
SN 2010jl was detected in a nearby star-forming galaxy UGC 5189A \citep{Newton10}, whose distance is measured to be $49$ Mpc. Observations from UV to NIR revealed it to be a luminous Type IIn SNe with bolometric luminosity of at least $3\times 10^{43}\ {\rm erg\ s^{-1}}$ \citep{Zhang12,Ofek14,Fransson14}. Various studies converge on the CSM to be massive, with estimates of the mass-loss rate ranging $4\times 10^{-2}$--$1\ M_\odot/{\rm yr}$ \citep{Zhang12,Moriya13, Ofek14, Fransson14,Chandra15}.

X-ray emission was detected from around 40 days after the explosion \citep{Chandra12}. Long-term X-ray observations by \cite{Chandra15} found an evolving column density from $N_H\sim 10^{24}\ {\rm cm^{-2}}$ to $10^{21}\ {\rm cm^{-2}}$ over a few years, and radio emission was first detected on day 570. This implies that the shock has been running through a dense CSM that efficiently absorbs X-ray and radio emission.

We compare our model to the bolometric and X-ray light curves of SN 2010jl. We adopt the parameters close to those estimated by previous works; ejecta with energy $1.6\times 10^{51}$ ergs for mass of $10M_\odot$, an outer ejecta slope of $n=7$, and a CSM mass loss rate of $0.13 M_\odot/{\rm yr}$. As an input of CLOUDY we adopt a CSM metallicity of $0.35$ solar, which is around the upper limit inferred from observations of the host galaxy \citep{Stoll11}. We compare the light curves only up to around day $200$. The bolometric light curve shows a break between day $200$ to $400$ \citep{Fransson14}, which can be because either the forward shock reached the outer edge of the dense CSM or the reverse shock reached the inner ejecta. In either case, our assumption of self-similarity in deriving the shock dynamics is inapplicable beyond this time.

\begin{figure*}
 \centering
\includegraphics[width=\linewidth]{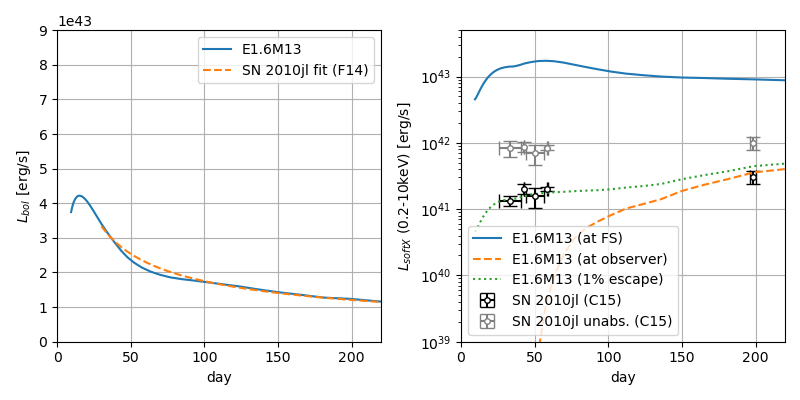}
\caption{Light curve model aimed to reproduce the observations of SN 2010jl. The left panel shows our bolometric light curve with a power-law fit of the observed light curve in \cite{Fransson14}. The right panel shows our unabsorbed and absorbed X-ray (0.2-10 keV) light curve with the data points taken from \cite{Chandra15}. 
We adopt the outer ejecta density power-law index of $n=7$ \citep{Chandra15} which roughly reproduces the decay rate of the bolometric light curve. We choose the explosion epoch to be JD 2,455,479, as chosen in \cite{Fransson14}.}
\label{fig:2010jl}
\end{figure*}

The result is shown in Figure \ref{fig:2010jl}. We find the bolometric light curve matches well with the best fit obtained from observations in \cite{Fransson14}. Both the FS and RS are radiative throughout the time range we consider, which is consistent with the analytical estimate by \cite{Fransson14}.

While the bolometric light curve is consistent, the X-ray luminosity in the right panel is not reproduced. Our predicted X-ray luminosity (orange dashed line) falls far below the observed one (black points). Furthermore, the X-ray luminosity at the forward shock front obtained from our model (solid line) overshoots the unabsorbed luminosity, calculated from spectral analysis of observational data (gray points) by \cite{Chandra15}.

These inconsistencies can be reconciled by abandoning the simplest assumption of the CSM to be spherical, and by e.g. introducing the existence of clumps or asphericity as mentioned in previous works \citep{Fransson14,Chandra15,Katsuda16}. To show this, we consider a case where 1\% of the emission from the shock front can escape, possibly due to the incomplete coverage of the CSM. The bolometric light curve in the left panel of Figure \ref{fig:2010jl} is hardly affected with this small solid angle.

\begin{figure}
     \centering
     \includegraphics[width=\linewidth]{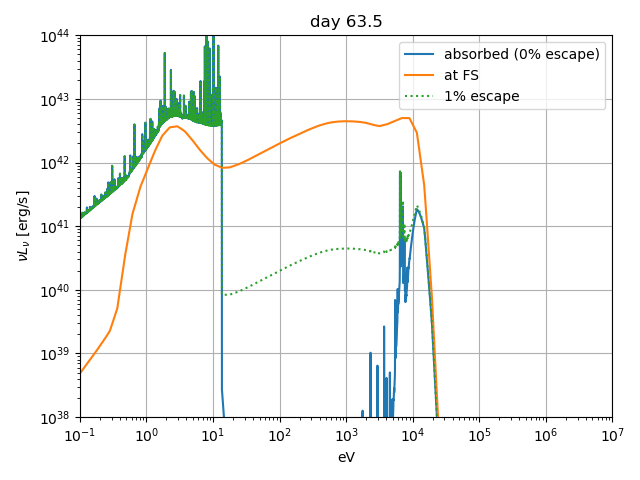}
     \caption{Spectrum at around day 60 of our model that reproduces the bolometric and X-ray light curves of SN 2010jl. Dashed and dotted lines show the escape fraction of $0$\% (spherical CSM) and $1$\% respectively.}
     \label{fig:2010jl_spec}
\end{figure}

The X-ray light curve in this case is shown as green dotted lines, which roughly matches with the observed X-ray luminosity up to day 200. With this 1\% escape fraction, the X-ray spectrum at around 60 days (Figure \ref{fig:2010jl_spec}) extends to about 10 keV, consistent with Chandra observations at a similar epoch \citep{Chandra15,Katsuda16}. Intriguingly, an order 1\% escape fraction is also inferred independently from the modelling of the radio emission \citep{Murase19}, though the radio data were taken much after the time range we consider (at around day 600). Nonetheless, SN 2010jl demonstrates that observations in both optical and X-rays would be an effective way for diagnosing the presence of asymmetry in the CSM.

\subsection{SN 2014C}
SN 2014C was first detected by the Lick Observatory Supernova Search \citep{Kim14}. It was originally classified as an SN Type Ib in a host galaxy NGC 7331, at a distance of $15$ Mpc. However, from $\sim 300$ days after explosion the source showed strong hydrogen emission lines analogous to Type IIn \citep{Milisavljevic15}, and became bright in X-rays \citep{Margutti17}. This transition from a free expansion to a strong interaction indicates that the progenitor exploded inside a shell-like massive CSM embedding a low-density cavity.

We compare our model with the X-ray observations of SN 2014C. However our dynamical modelling, which assumes an extended CSM with a power-law density profile, cannot be applied to a shell-like CSM argued for SN 2014C. Instead, we focus on one epoch, at day 400, and construct an X-ray spectrum at this epoch.

The X-ray spectrum at this epoch obtained by \cite{Margutti17} shows a clean bremsstrahlung emission with a characteristic temperature of 18 keV. From this spectrum \cite{Margutti17} deduce that the X-ray emission mainly originates from the optically thin FS region, which is close to adiabatic and moving at a velocity of $V_{\rm sh}\sim 4000\ {\rm km\ s^{-1}}$. The shock radius at this epoch is estimated to be $6.4\times 10^{16}$ cm from VLBI imaging \citep{Bietenholz18}. We adopt these values and use our model with only the FS component to obtain the spectrum, as the RS component is argued to be negligible \citep{Margutti17}. We vary the downstream number density $n_{\rm down}$ as a parameter. We put a CSM of solar metallicity \citep{Milisavljevic15} with a flat density profile with number density $n_{\rm down}/4$. The 4 in the denominator is the compression ratio for $\gamma=5/3$, as this shock is close to adiabatic. The width of the CSM is determined so that the hydrogen column density is $3\times 10^{22}\ {\rm cm^{-2}}$, as obtained from X-ray spectral fitting. For this case the advection time $r_{\rm sh}/V_{\rm sh}\sim 5$ years is much bigger than 400 days. Thus we only calculate with our model up to 400 days, and obtain $L_{\rm \nu}$ as
\begin{eqnarray}
L_\nu = 4\pi r_{\rm sh}^2 \frac{r_{\rm sh}}{4} \frac{e_{\rm esc,\nu}(t=400\ {\rm day})}{400\ {\rm day}}.
\end{eqnarray}
This time we do not include the factor 2 accounting for direction, as we do not consider the reprocessing by the (presumably optically thin) RS region. We also neglect the effect of diffusion since the diffusion timescale is negligible at this epoch.

\begin{figure*}
 \centering
\includegraphics[width=0.75\linewidth]{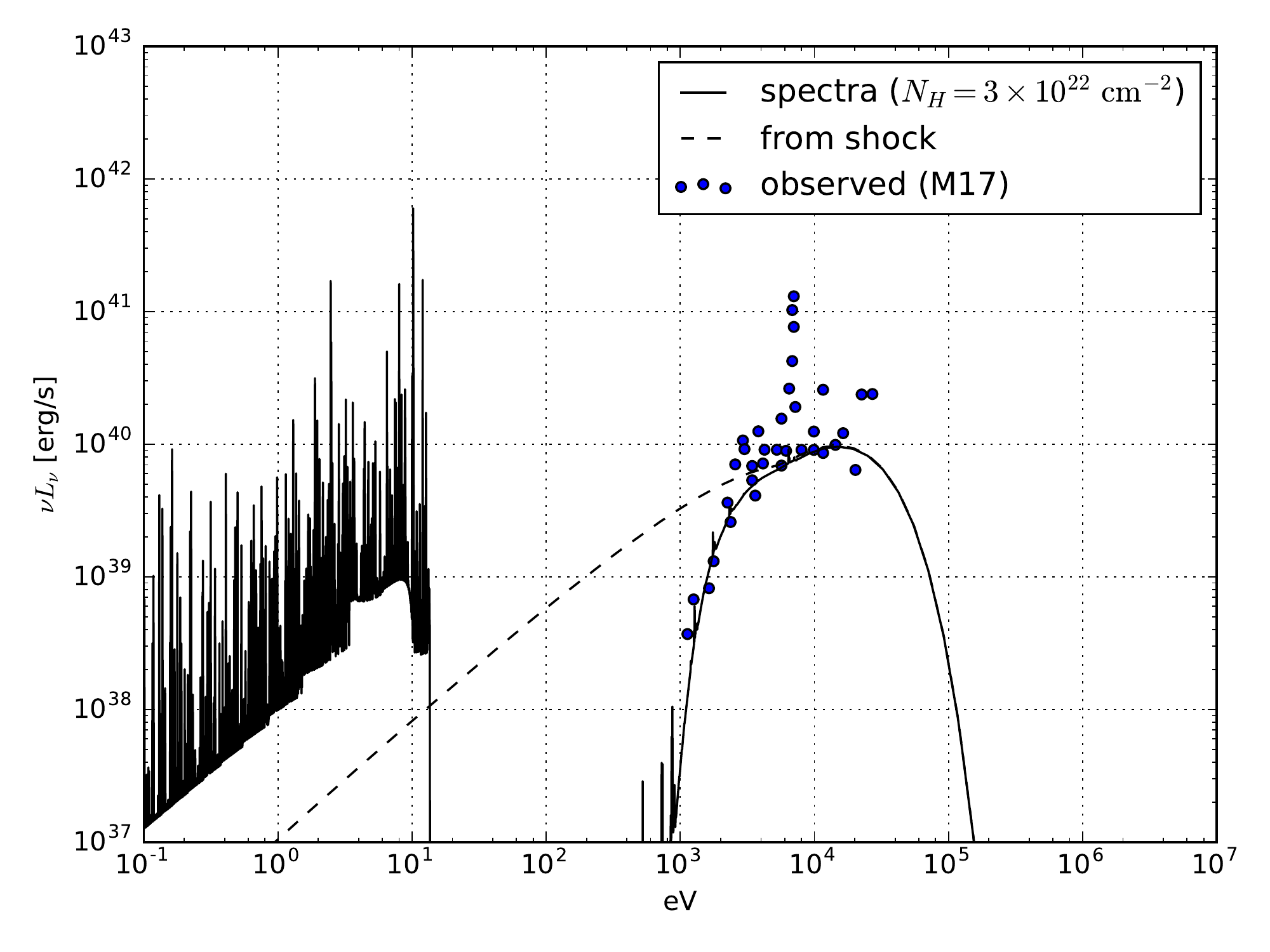}
\caption{Spectrum of our SN 2014C model at 400 days after explosion. The blue points show the X-ray observations from \cite{Margutti17}. We adopt a shock radius of $6.4\times 10^{16}$ cm, shock velocity of $4000$ km/s, and downstream number density $1.4\times 10^6\ {\rm cm}^{-3}$.}
\label{fig:2014C}
\end{figure*}

We compare our spectrum (solid lines) and the observed spectrum by \cite{Margutti17} (blue points) in Figure \ref{fig:2014C}. Our spectrum, with a downstream density of $n_{\rm down}=1.4\times 10^6\ {\rm cm^{-3}}$, reasonably matches the bremsstrahlung component in \cite{Margutti17}. The prominent H$\alpha$ line observed in \cite{Milisavljevic15} is also seen in our spectrum, although a detailed comparison is subject to uncertainties in the profile and geometry of the CSM, and is beyond the scope of this work.
One caveat is that the strong excess at 6.7--6.9 keV, which \cite{Margutti17} attributes to K$\alpha$ transitions of H- and He-like ions of iron, is not seen in our model. This excess is seen in SN 2006jd as well, and may be explained by an enhancement of iron compared to solar, or co-existence of a cooler component from e.g. RS region or clumps (see \citealt{Chandra12_2006jd,Margutti17} for discussion).

\section{Caveats}
\label{sec:caveats}
Our emission model ignores some physics that may have impact on the emission. As this work is the first step to derive detailed optical to X-ray spectra, incorporation of these are to be dealt with in future work. In this section we raise the caveats of our work, and discuss qualitatively in what cases these may become important.

\subsection{Opacity in the Shocked Region}
As mentioned in Section \ref{sec:formulation} we have included only free-free processes as the emission/absorption mechanism of photons in the shocked region. Although this may be a good approximation at low-metallicity environments, there are additional contribution from bound-free (and bound-bound) processes that can be important at metallicities around and beyond solar. As seen in Figure \ref{fig:Te_evolution} the plasma temperature in the shocked region can drop down to a temperature $\sim 10^4$--$10^5$ K. At this temperature metals are not fully ionized and thus can efficiently convert soft X-rays to UV and optical radiation\footnote{We note that these processes are taken into account in the radiation transfer calculation through the CSM done with CLOUDY, as observed in Fig \ref{fig:spectrum}.}.

We defer the precise modelling of this to future work, but the qualitative effect of this may be seen by enhancing the opacity by a factor $\eta = [1+(\kappa_{\rm bb}+\kappa_{\rm bf})/\kappa_{\rm ff}]$, where $\kappa_{\rm bb}$, $\kappa_{\rm bf}$ and $\kappa_{\rm ff}$ are the bound-bound, bound-free and free-free opacities \citep{Svirski12}. We crudely assume that $\eta$ is independent of frequency and temperature, i.e. the free-free emissivity and absorption coefficient (equations \ref{eq:j_ff} and \ref{eq:alpha_ff}) are multiplied by $\eta$, with everything else unchanged. We compare our results (assuming $\eta=1$) for the E1M1 model to the case where $\eta=10$. 

\begin{figure*}[t]
 \begin{tabular}{cc}
 \begin{minipage}{0.5\hsize}
 \centering
\includegraphics[width=\linewidth]{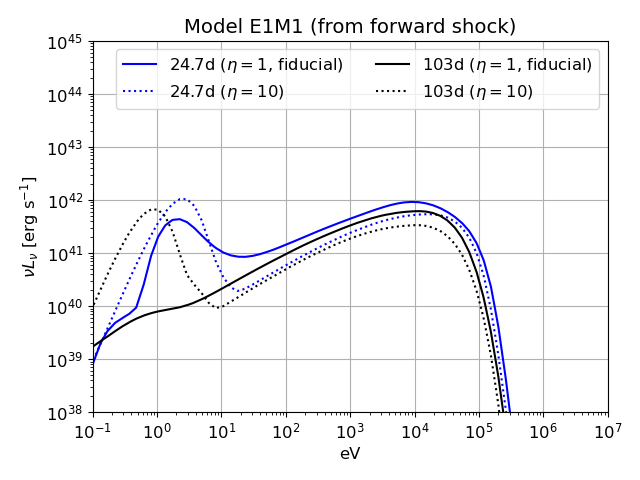} 
\end{minipage}

\begin{minipage}{0.5\hsize}
 \centering
\includegraphics[width=\linewidth]{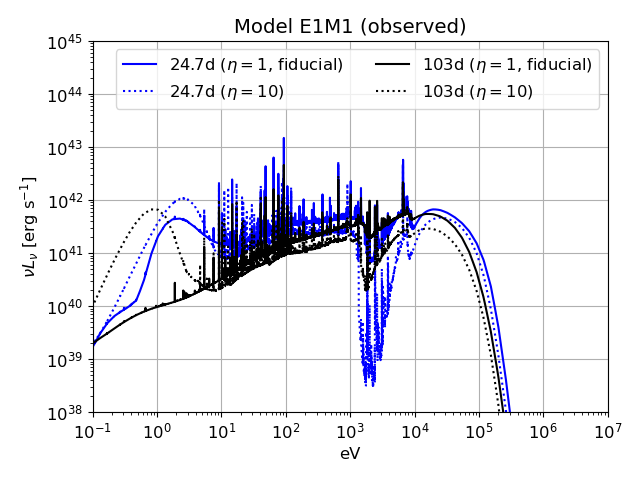}
\end{minipage}\\

\end{tabular}
\caption{Spectrum $\nu L_\nu$ for the model E1M1 with and without the opacity enhancement factor $\eta$. The left panel shows the spectrum of radiation escaping from the FS region, and the right panel shows that after passing through the CSM.}
\label{fig:opac10x}
\end{figure*}

The comparison can be seen in Figure \ref{fig:opac10x}. The left panel shows the spectrum at the shock radius, and the right panel shows the spectrum that is reprocessed through the CSM. We generally see that for a larger $\eta$ there is enhancement of the optical component at the cost of reducing the X-ray component. The difference is most clearly seen at the late epoch. Thus we conclude that inclusion of a more realistic bound-free opacity would be important for characterising the emission in the optical and soft X-rays, although this is included in the radiation transfer through the unshocked CSM. However, we should note that the hard X-ray spectrum should be close to the $\eta=1$ one, as plasma reaching temperature hotter than $\sim10$ keV is fully ionized and free-free emission should dominate.

\subsection{Feedback from Radiative Cooling}
\label{sec:caveat_rad_cool}
Another caveat of our work is that we neglect the effect of radiative cooling onto the dynamics. The effect of this can be summarized into two aspects.

One is in the shock propagation. In this work we have adopted the self-similar model of shock evolution by \cite{Chevalier82}, which assumes the shock downstream is adiabatic. We observe in some cases the shock becoming radiative, where solving the momentum equation assuming a thin shell \citep{Chevalier03} is more appropriate. In this case we obtain the time evolution of the shock radius as $r_{\rm sh}=[8\pi v_w g^n/(n-4)(n-3)\dot{M}]^{1/(n-2)}t^{(n-3)/(n-2)}$,
which differs from equation (\ref{eq:rsh_ss}) by a factor $[\alpha(n-4)(n-3)/2]^{1/(n-2)}$. In the case of our model ($\alpha=0.0595, n=10$) this is merely a $\sim 3\%$ difference, which affects the bolometric light curve ($\propto r_{\rm sh}^2\rho_{\rm w} v_{\rm sh}^3$ for a radiative case) by $\lesssim 10\%$. We conclude that refinement of the shock dynamics will only have a small impact on our results.

The other point, perhaps more important, is that we neglect the density gradient in the shocked region for simplicity. The radiative processes adopted in our model depend on the plasma density, and thus cooling and heating will accelerate if this compression is taken into account. This can be partially taken into account by adopting a lower adiabatic index close to 1, which enhances the downstream density. 

\begin{figure*}[t]
 \begin{tabular}{cc}
 \begin{minipage}{0.5\hsize}
 \centering
\includegraphics[width=\linewidth]{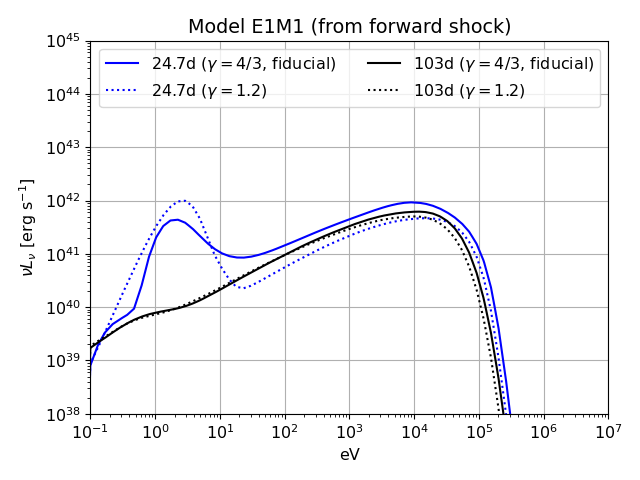} 
\end{minipage}

\begin{minipage}{0.5\hsize}
 \centering
\includegraphics[width=\linewidth]{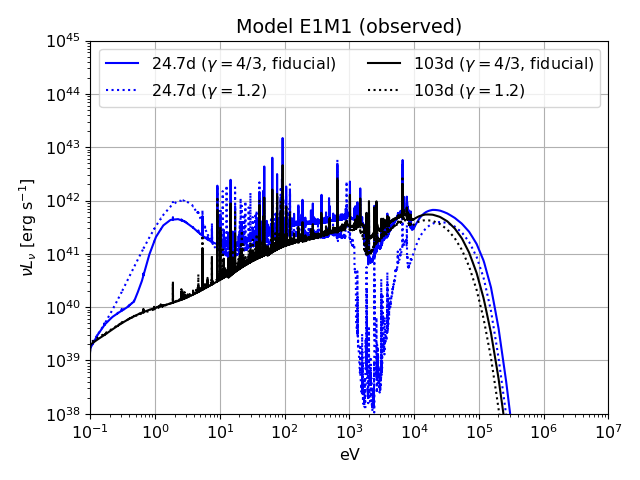}
\end{minipage}\\

\end{tabular}
\caption{Same as Figure \ref{fig:opac10x}, but comparison by the adiabatic index $\gamma$. We compare $\gamma=4/3$ used in our models and a lower $\gamma=1.2$.}
\label{fig:gamma1.2}
\end{figure*}

To see the effects of lowering $\gamma$ we compare our E1M1 model (assuming $\gamma=4/3$) with a result assuming $\gamma=1.2$. With this low $\gamma$ the downstream density is enhanced due to the higher compression ratio (factor $11/7$), and slightly more due to the lower value of $\alpha$ by order 10\%. Figure \ref{fig:gamma1.2} shows the comparison of these two values of $\gamma$ at the same epochs. In the early phase, the density enhancement helps convert more UV/X-ray photons to optical, as shown in the left panel. This can have a large impact on the absorption of soft X-rays around 1 keV, as shown in the right panel for day 24.7. In the late phase, the density has sufficiently dropped and the resulting spectra become almost the same.

\subsection{Collisionless Relaxation}
We have considered Coulomb collision as a relaxation process for protons and electrons. However it is known that for interactions of the ejecta and CSM, the FS turns collisionless around breakout \citep{Katz11, Murase11}. As a result collisionless relaxation can occur, which can modify the relaxation timescale of the plasma. Furthermore, a non-negligible fraction of the post-shock energy may be used to accelerate electrons and protons to relativistic energies through the diffusive shock acceleration mechanism (e.g. \citealt{Blandford87}). 

The latter are proposed to give rise to various emissions in both photons and neutrinos \citep{Murase11,Katz11,Murase14,Zirakashvili16,Petropoulou17,Murase18,Murase19}. The particles and their emissions also interact with optical/X-ray photons through e.g. inverse Compton processes and two-photon annihilation processes. Detailed modelling of the spectrum like done in this work may thus be important for deriving the non-thermal emission from these relativistic particles. In future work we plan to apply our current emission model to calculate the non-thermal emission of interaction-powered transients. We note that these photons are expected to be prominent in radio and gamma-rays, and we expect inclusion of this will not greatly modify our present results which cover from optical to X-rays.

\section{Conclusion}
\label{sec:conclusion}

\begin{figure*}
 \centering
\includegraphics[width=0.9\linewidth]{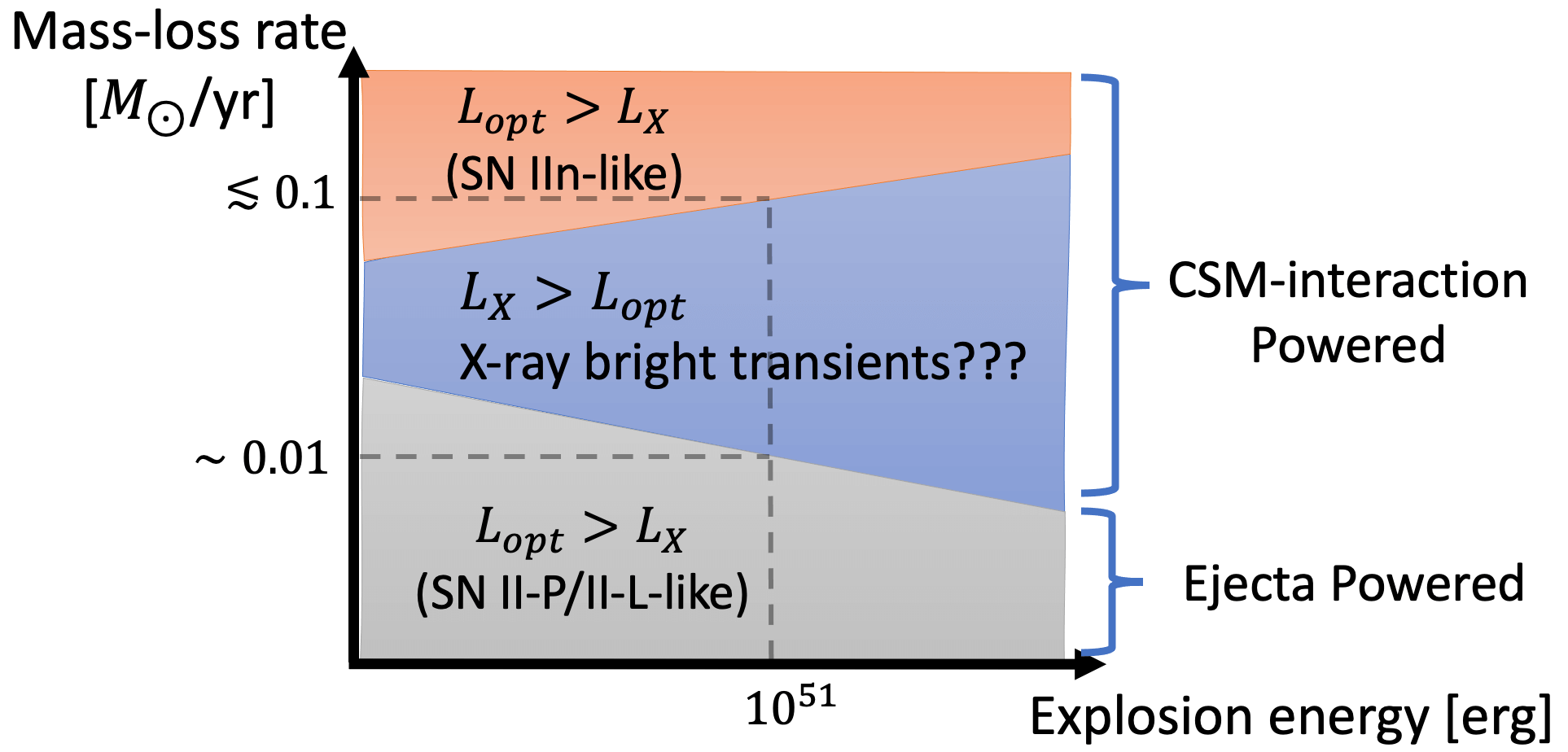}
\caption{A summary figure of this work, showing the possible diversity of the multi-wavelength properties of (hydrogen-rich) interaction-powered transients, based on the ejecta energy and mass loss rate of the CSM. The borders $\dot{M}$ of adjacent regions in the y-axis are rough values implied from our results for $v_w=100\ {\rm km\ s^{-1}}$, and should scale as $\dot{M}\propto v_w$.}
\label{fig:summary}
\end{figure*}
In this work we constructed for the first time a theoretical model for obtaining a multi-wavelength spectra of interaction-powered SNe. Our model takes into account the important physical processes in the shocked region, such as the collisional relaxation of electrons and protons, and the comptonization/absorption of bremsstrahlung radiation. 

We focused on the parameter space relevant to Type IIn SNe, and found that the mass-loss rate has great impact on shaping the spectra and the relative brightness of X-rays compared with the optical/UV. In agreement to previous studies, we find that for high mass-loss rates X-ray emission is largely suppressed due to reprocessing in the shocked region and the CSM, and the transient becomes bright in optical/UV. On the other hand our model extends previous works by considering a lower mass-loss rate, and predicts that these events can become luminous X-ray sources, with luminosity $10^{42}$--$10^{44}\ {\rm erg\ s^{-1}}$ for up to 100 days. We summarize our results in Figure \ref{fig:summary}, with quantitative values roughly estimated from the results of our work. This paradigm would be testable by current/future all-sky surveys in the X-ray wavelength.

We furthermore tested our model with observations of SN 2010jl and SN 2014C. The comparison with SN 2010jl shows that asymmetric structure in the CSM is required to explain simultaneously the optical and X-ray light curves, in agreement with previous studies \citep{Fransson14,Chandra15}. For SN 2014C the X-ray and H$\alpha$ emission seen in the late phase are roughly reproduced.

Our model only requires the shock radius, shock velocity, and the upstream density as input to calculate the emitted spectrum. These parameters can be obtained not only from the self-similar solutions we adopted, but also by utilizing hydrodynamical simulations. Coupling our model with hydrodynamical simulations will greatly enhance the capability of our model, because we can relax various restrictions of the ejecta and CSM. We plan to explore such possibility in future work.

\acknowledgements
The authors thank Kohta Murase for many valuable comments throughout the course of this work, and thank Satoru Katsuda for important comments on SN 2010jl. DT is supported by the Advanced Leading Graduate Course for Photon Science (ALPS) at the University of Tokyo. This work is also supported by JSPS KAKENHI Grant Numbers JP19J21578, JP16H06341, JP20K04010, JP20H01904, JP20H05639, MEXT, Japan.

\appendix
\section{Shock dynamics}
\label{sec:dynamics}
The dynamics of SN ejecta colliding with a CSM of wind profile ($\rho=\dot{M}/4\pi v_wr^{2}$) was first obtained as a self-similar solution in \cite{Chevalier82}. This is applicable while the swept up mass is much smaller than the ejecta mass. The SN ejecta is assumed to be cold (i.e. negligible thermal pressure) and have a density profile with function of radius and time as \citep{Matzner99}
\begin{eqnarray}
\rho_{\rm ej} (r,\mathcal{T})
&=& \left\{ \begin{array}{ll}
\mathcal{T}^{-3}\left[r/(g\mathcal{T})\right]^{-n} & (r/\mathcal{T} > v_t),\\
\mathcal{T}^{-3}(v_t/g)^{-n} \left[r/(\mathcal{T}v_t)\right]^{-\delta}  & (r/\mathcal{T} < v_t),
\end{array}\right.
\label{eq:rho_ej}
\end{eqnarray}
where $\mathcal{T}$ is the elapsed time since explosion introduced in Section \ref{sec:twozone}, and $g, v_t$ are given by the ejecta mass $M_{\rm ej}$ and energy $E_{\rm ej}$ as
\begin{eqnarray}\label{eq:coeff_ej}
g &=& \left\{\frac{1}{4\pi(n-\delta)} \frac{[2(5-\delta)(n-5)E_{\rm ej}]^{(n-3)/2}}{[(3-\delta)(n-3)M_{\rm ej}]^{(n-5)/2}}\right\}^{1/n},
\label{eq:ejecta_g}\\
v_t &=& \left[\frac{2(5-\delta)(n-5)E_{\rm ej}}{(3-\delta)(n-3)M_{\rm ej}}\right]^{1/2}.\label{eq:v_t}
\end{eqnarray}
The outer ejecta have a steep density profile with $n=7$--$12$. In this case the radius at the contact discontinuity follows a power-law time evolution \citep{Chevalier82}
\begin{eqnarray}
    r_{\rm sh} &=& \left(\frac{4\pi v_w\alpha g^n}{\dot{M}}\right)^{1/(n-2)}\mathcal{T}^{(n-3)/(n-2)},
    \label{eq:rsh_ss}
\end{eqnarray}
where $\alpha$ is a parameter obtained numerically and depends on $n$ and the adiabatic index $\gamma$. The factor $\alpha^{1/(n-2)}$ is of order unity in most cases. The velocities of the FS and RS upstreams in the lab frame are from the Rankine-Hugoniot relations \citep{Tsuna19}
\begin{eqnarray}
    v_{\rm fs} &=& \frac{\gamma+1}{2}v_{\rm fs, down} = \frac{\gamma+1}{2}\frac{n-3}{n-2}\frac{r_{\rm sh}}{\mathcal{T}}, \\
    v_{\rm rs} &=& \frac{n-3}{n-2} \left[1-\frac{2}{\gamma+1}\frac{1}{n-2}\right]^{-1}v_{\rm rs, down} = \left[1-\frac{2}{\gamma+1}\frac{1}{n-2}\right]^{-1}\left(\frac{n-3}{n-2}\right)^2\frac{r_{\rm sh}}{\mathcal{T}},
\end{eqnarray}
where we assumed a strong shock and adopted the thin-shell approximation ($v_{\rm fs, down}\approx v_{\rm rs, down}\approx dr_{\rm sh}/d\mathcal{T}$).

In this work we adopt $n=10$, $\delta=1$, $\gamma=4/3$, and $M_{\rm ej}=10M_\odot$ unless otherwise noted. By numerically obtaining the self-similar solutions we get $\alpha\approx 0.0595$, and
\begin{eqnarray}
r_{\rm sh}&\approx& 5.7\times 10^{14}{\rm cm}\ \left(\frac{\mathcal{T}}{10\ {\rm day}}\right)^{7/8} \left(\frac{M_{\rm ej}}{10M_\odot}\right)^{-5/16}
\left(\frac{E_{\rm ej}}{10^{51}{\rm erg}}\right)^{7/16} \left[\frac{\dot{M}/v_w}{(\dot{M}/v_w)_*}\right]^{-1/8},
\end{eqnarray}
where we adopt a fiducial mass-loss rate for normalization $(\dot{M}/v_w)_*\equiv 10^{-4}M_\odot\,{\rm yr}^{-1}/({\rm km\ s}^{-1})$, corresponding to $10^{-2}M_\odot\,{\rm yr}^{-1}$ for a wind velocity of $v_w=100\ {\rm km\ s}^{-1}$.
For a wind CSM ($\rho\propto r^{-2}$), the downstreams of the two shocks typically have completely different densities \citep{Chevalier82}. The FS downstream density $\rho_{\rm fs}$ is $(\gamma+1)/(\gamma-1)=7$ times the upstream density,
\begin{eqnarray}
    \rho_{\rm fs}&\approx&1.1\times 10^{-13}\ {\rm g\ cm^{-3}} \ \left(\frac{\mathcal{T}}{10\ {\rm day}}\right)^{-7/4}\left(\frac{M_{\rm ej}}{10M_\odot}\right)^{5/8}
\left(\frac{E_{\rm ej}}{10^{51}{\rm erg}}\right)^{-7/8} \left[\frac{\dot{M}/v_w}{(\dot{M}/v_w)_*}\right]^{5/4}.
\end{eqnarray}
From the Rankine-Hugoniot relations and adopting the thin-shell assumption, the density ratio at the RS and FS downstreams is $\rho_{\rm rs}/\rho_{\rm fs}\approx \alpha^{-1}$ \citep{Tsuna19}, which is $\approx 17$ for our parameters.

At a sufficiently late time, the outer ejecta are completely swept up by the RS, after which the self-similarity breaks down. For the above parameters, this occurs at 
\begin{equation}
\mathcal{T}\approx 960\ {\rm days}\left(\frac{M_{\rm ej}}{10M_\odot}\right)^{3/2}\left(\frac{E_{\rm ej}}{10^{51}{\rm erg}}\right)^{-1/2}\left[\frac{\dot{M}/v_w}{(\dot{M}/v_w)_*}\right]^{-1}.
\end{equation}

We use these formulations to obtain the shock radius $r_{\rm sh}$, shock velocities ($v_{\rm fs}, v_{\rm rs}$), and downstream densities ($\rho_{\rm fs},\rho_{\rm rs}$) at a given epoch \{$\mathcal{T}_0,\mathcal{T}_1,\cdots$\}. These values will be given as inputs to calculate the emitted radiation spectrum.

\section{Ultra-fast Cooling Case}
\label{sec:Ultra-fast cooling}
When solving the electron cooling in the early epochs, the free-free cooling timescale $t_{\rm cool}$ in the RS region (equation \ref{eq:t_cool_rs}) became very small, in extreme cases less than 10 seconds at 10 days after explosion. This resulted in difficulties in accurately obtaining the radiation energy densities from equations (\ref{eq:erad}) and (\ref{eq:eesc}), because to compute these equations accurately one has to keep records of $j_{\nu, \rm ff}$ with resolution of $t_{\rm cool}$ for the past $t_{\rm diff}$ (which increases with $\dot{M}$). The need to allocate this amount of memory made it impossible with our computational resource to numerically obtain the radiation energy densities.

Fortunately though, this heavy computation should not be required. In this regime, the free-free processes dominate over any other processes, and the timescale for the plasma to approach thermal equilibrium with the radiation is much shorter than the diffusion or dynamical timescales. Thus we can safely assume that the plasma is always in thermal equilibrium with radiation, with blackbody temperature given by the total energy density $\int d\nu e_{\rm rad, \nu}$ available at that time in the RS region. Motivated by this, we analytically obtain the approximate $e_{\rm rad,\nu}$ and $e_{\rm esc, \nu}$ at the end $t=t_{\rm dyn}$ in the following way.

The thermal energy initially carried by the downstream $e_{\rm th, rs}\equiv (3/2)n_ik_{\rm B}(T_{p,\rm rs}+T_{e,\rm rs})$ is instantaneously converted to radiation within $t=t_{\rm cool}$, and is radiated away within $t_{\rm diff}+t_{\rm cool}\approx t_{\rm diff}$. There are also photons injected from the FS region at rate $j^{RS}_{\rm \nu, in}$ [erg/s/cm$^3$/Hz]; what is injected from $0<t<t_{\rm dyn}-t_{\rm diff}$ will be radiated away while the contribution over the last $t_{\rm diff}$ will remain at $t=t_{\rm dyn}$. 


The bolometric values $\int d\nu e_{\rm rad,\nu}$ and $\int d\nu e_{\rm esc,\nu}$ are thus
\begin{eqnarray}
    \int d\nu\ e_{\rm rad,\nu} &=& 
    t_{\rm diff}\int d\nu j^{RS}_{\nu,\rm in} \label{eq:eradnubol_appB}\\
    \int d\nu\ e_{\rm esc,\nu} &=& e_{\rm th}+(t_{\rm dyn}-t_{\rm diff})\int d\nu j^{RS}_{\nu,\rm in}.
    \label{eq:eescnubol_appB}
\end{eqnarray}
The spectrum $e_{\rm rad,\nu}$ is a blackbody spectrum with total energy density given in equation (\ref{eq:eradnubol_appB}). On the other hand, the spectrum $e_{\rm esc,\nu}$ will slightly differ for (i) $0.5t_{\rm dyn}<t_{\rm diff}<t_{\rm dyn}$ and (ii) $t_{\rm diff}<0.5t_{\rm dyn}$. 
For case (i), $e_{\rm esc,\nu}$ is a single-component blackbody spectrum with total energy density given in equation (\ref{eq:eescnubol_appB}). For case (ii), $e_{\rm esc, \nu}$ is a sum of two components, the first being a blackbody spectrum with total energy density $e_{\rm th}+t_{\rm diff}\int d\nu j^{RS}_{\nu,\rm in}$, and the second being a blackbody spectrum with total energy density $t_{\rm diff}\int d\nu j^{RS}_{\nu,\rm in}$, multiplied by $(t_{\rm dyn}-2t_{\rm diff})/t_{\rm diff}$. This is because for case (ii), there will be a time range $t_{\rm diff}<t<t_{\rm dyn}-t_{\rm diff}$ where the total radiation energy density is limited to $\approx t_{\rm diff}\int d\nu j^{RS}_{\nu,\rm in}$.

\bibliographystyle{apj}
\bibliography{references}

\end{document}